\documentclass[print]{aa}

\usepackage{graphicx}
\usepackage{CJK}
\usepackage{enumitem}
\usepackage[varg]{txfonts}
\usepackage{amsmath}
\usepackage{amssymb}
\usepackage{booktabs}

\begin{document}

\begin{CJK*}{UTF8}{gbsn}

\title{Red supergiant stars in the Large Magellanic Cloud. \uppercase\expandafter{\romannumeral2}.\\ Infrared properties and mid-infrared variability
}
\titlerunning{RSGs in the LMC \uppercase\expandafter{\romannumeral2}. IR properties and MIR variability}

\author{
Ming Yang (杨明) \inst{1} \and Alceste Z. Bonanos \inst{1} \and Bi-Wei Jiang (姜碧沩) \inst{2} \and Jian Gao (高健) \inst{2} \and Meng-Yao Xue (薛梦瑶) \inst{3} \and Shu Wang (王舒) \inst{4} \and Man I Lam (林敏仪) \inst{5} \and Zoi T. Spetsieri \inst{1} \and Yi Ren (任逸) \inst{2} \and Panagiotis Gavras \inst{1}
}
\authorrunning{Yang, Bonanos \& Jiang et al.}

\institute{
Institute for Astronomy, Astrophysics, Space Applications \& Remote Sensing, National Observatory of Athens, Vas. Pavlou and I. Metaxa, Penteli 15236, Greece\\
                \email{myang@noa.gr, bonanos@noa.gr} \and
Department of Astronomy, Beijing Normal University, Beijing 100875, P. R. China \and
International Centre for Radio Astronomy Research, Curtin University, Bentley, WA 6102, Australia \and
Kavli Institute for Astronomy and Astrophysics, Peking University, Beijing 100871, P. R. China \and
Key Laboratory for Research in Galaxies and Cosmology, Shanghai Astronomical Observatory, Chinese Academy of Sciences, 80 Nandan Road, Shanghai 200030, P. R. China
                }

\abstract{
The characteristics of infrared properties and mid-infrared (MIR) variability of red supergiant (RSG) stars in the Large Magellanic Cloud (LMC) are analyzed based on 12 bands of near-infrared (NIR) to MIR co-added data from 2MASS, Spitzer and WISE, and $\sim$6.6 years of MIR time-series data collected by the ALLWISE and NEOWISE-R projects. 773 RSGs candidates were compiled from the literature and verified by using the color-magnitude diagram (CMD), spectral energy distribution (SED) and MIR variability. About 15\% of valid targets in the $IRAC1-IRAC2$/$IRAC2-IRAC3$ diagram may show polycyclic aromatic hydrocarbon (PAH) emission. We show that arbitrary dereddening Q parameters related to the IRAC4, S9W, WISE3, WISE4, and MIPS24 bands could be constructed based on a precise measurement of MIR interstellar extinction law. Several peculiar outliers in our sample are discussed, in which one outlier might be a RSG right before the explosion or an extreme asymptotic giant branch (AGB) star in the very late evolutionary stage based on the MIR spectrum and photometry. There are 744 identified RSGs in the final sample having both the WISE1- and WISE2-band time-series data. The results show that the MIR variability is increasing along with the increasing of brightness. There is a relatively tight correlation between the MIR variability, mass loss rate (MLR; in terms of $K_S-WISE3$ color), and the warm dust or continuum (in terms of WISE4 magnitude/flux), where the MIR variability is evident for the targets with $K_S-WISE3>1.0~mag$ and $WISE4<6.5~mag$, while the rest of the targets show much smaller MIR variability. The MIR variability is also correlated with the MLR for which targets with larger variability also show larger MLR with an approximate upper limit of $-6.1~M_\sun/yr^{-1}$. Both the variability and the luminosity may be important for the MLR since the WISE4-band flux is increasing exponentially along with the degeneracy of luminosity and variability. The identified RSG sample has been compared with the theoretical evolutionary models and shown that the discrepancy between observation and evolutionary models can be mitigated by considering both variability and extinction.
}


\maketitle

\section{Introduction}

Massive stars with a mass range roughly between $\sim8-30~M_\sun$ spend a fraction of their life time as red supergiant (RSG) stars located in the luminous cool region of the Hertzsprung-Russell (H-R) diagram. As the descendants of blue H-burning OB-type massive stars, RSGs are thought to horizontally evolve across the H-R diagram and become either Type-II supernovae (SNe) or blue supergiant and/or Wolf-Rayet (W-R) stars depending on their initial masses, chemical compositions and mass-loss rates (MLR) \citep{Eggenberger2002, Smartt2004, Smartt2009, Humphreys2010, Meynet2011, Georgy2012}. As He-burning, evolved, Population I stars, their unique and extreme physical properties represent an important phase of massive stellar evolution. Typically, RSGs have relatively young ages of $\sim8-20~Myr$, low effective temperatures ($T_{eff}$) ranging from $\sim3500-4500~K$, high luminosities of 4000 to 400000 $L_\sun$, and radii up to 1500 $R_\sun$ \citep{Humphreys1979c, Levesque2005, Massey2008, Levesque2009, Wittkowski2012, Ekstrom2013, Massey2013}. 

It is a challenge to accurately determine the physical properties of RSGs due to their large size and extended atmospheres, cool $T_{eff}$, high extinction or reddening, and semi-regular variability. However, despite the intrinsic difficulties in quantifying the physical parameters, the combination of high luminosity and emission peak in the near infrared (NIR) bands also allows them to be observed at large distances or highly obscured environments at these wavelengths compared to the optical bands \citep{Davies2007, Clark2009, Negueruela2010, Britavskiy2014, Britavskiy2015, Williams2015, Messineo2016, Davies2017}.

Besides the physical limitations, a fair and representative sample volume, covering both bright and faint ends of magnitude and different metallicities, is also important to understand the formation and evolution of RSGs. Therefore, as the first steps of the cosmic distance ladder, the Magellanic Clouds (MCs) play a crucial role not only in the extragalactic distance scale, but also in the chemical evolution and metallicity distribution, since the metallicities of the MCs are about half ($Z=0.009$) and 1/4 ($Z=0.005$) of the Milky Way, respectively \citep{Keller2006}. However, until several recent large-scale photometric and spectroscopic surveys, the studies of RSGs in the Milky Way and the MCs had been limited to the bright population \citep{Humphreys1979a, Humphreys1979b, Elias1985, Oestreicher1997, Massey2003, Levesque2006}. 

The most recent efforts to characterize the nature of RSGs in the MCs are approaching the problem from both optical and infrared (IR) directions. In the early 2000s, \citet{Massey2002} conducted a UBVR CCD survey of the MCs to obtain well-calibrated data on the brighter, massive stars. Following the photometric survey, they also confirmed that most of their candidates were RSGs by obtaining optical spectra for a significant part of the sample \citep{Massey2003}. \citet{Neugent2012} performed further analysis based on the 2MASS photometry and optical spectra and greatly increased the number of RSGs in the Large Magellanic Cloud (LMC). Another dedicated optical spectroscopic survey by \citet{Gonzalez2015} also approached the faint end of the RSGs in the MCs to build a more representative dataset. In the infrared bands, \citet{Bonanos2009, Bonanos2010} explored the mid-infrared (MIR) properties of massive stars in the MCs and derived new MIR criteria to identify RSGs, which were used by \citet{Britavskiy2014} and \citet{Britavskiy2015} to identify new RSGs in Sextans A, IC 1613 and a few more dwarf irregulars, and \citet{Williams2015} who applied it to M83. Using RSG candidates taken from the literature, \citet{Yang2011} (hereafter Paper \uppercase\expandafter{\romannumeral1}) and \citet{Yang2012} also investigated the NIR to MIR properties of RSGs in the MCs and derived the IR period-luminosity (P-L) relation of RSGs based on the optical time-series data from the All Sky Automated Survey (ASAS; \citealt{Pojmanski2002}) and the MAssive Compact Halo Objects (MACHO; \citealt{Alcock1997}) projects. The increasing volume of data from both ground and space telescopes and the expanded sample of RSGs in the MCs, now give us an opportunity to complement our knowledge of the IR properties and the unexplored MIR time-domain information of RSGs, which hopefully will be further achieved with the JWST.

In this paper, we present the IR properties and MIR variability analysis of the largest sample of RSGs in the LMC up to now. The sample selection and photometry are presented in \textsection2 and \textsection3. The data analysis is described in \textsection4. The discussion and summary are given in \textsection5 and \textsection6, respectively. 

\section{Sample selection}

Our sample is compiled from various works in the literature in which RSGs have been identified by either photometry or spectroscopy. In 2011 (Paper \uppercase\expandafter{\romannumeral1),} we studied the P-L relation of RSGs in the LMC by using a sample of 189 RSGs verified by their brightness and color indexes in several NIR and MIR bands, that is, Two Micron All Sky Survey (2MASS)/$JHK_{S}$ \citep{Skrutskie2006}, the Spitzer/IRAC and MIPS bands \citep{Werner2004}. This sample also included previous photometric selected RSGs in the LMC by using nonuniform criteria \citep{Feast1980, Pierce2000, Kastner2008}. The majority of targets ($\sim70\%$) in this sample were also spectroscopically confirmed by \citet{Massey2003} (hereafter M03). With a similar methodology, the sample of RSGs in the LMC was largely expanded by \citet{Neugent2012} (hereafter N12) who evaluated the evolutionary modeling of Yellow Supergiant stars (YSGs) and RSGs based on the new Geneva evolutionary models. The sample of 505 RSGs was mainly selected by using magnitude and color, along with the constrains on the proper-motion and the systemic radial velocity of LMC to separate out foreground dwarfs. Recently, \citet{Gonzalez2015} (hereafter G15) conducted a pilot program aimed at exploring the fainter end of RSGs and extrapolated their behavior to other environments by building a more representative sample in the MCs. As previous works did, they used near infrared selection criteria, spectroscopically classified the targets and derived their radial velocities to confirm the membership in the clouds. With more than two hundred newly discovered RSGs in the MCs (123 in LMC), they study the physical properties of about 500 RSGs by using NIR/MIR photometry and optical spectroscopy. More recently, \citet{Jones2017} (hereafter J17) classified nearly 800 point sources in the LMC, which were observed by the infrared spectrograph (IRS) on board the Spitzer Space Telescope, into different types according to their infrared spectral features, continuum and spectral energy distribution shape, bolometric luminosity, cluster membership and variability information, by using a decision-tree classification method. 

\begin{figure}
\includegraphics[trim=3cm 0cm 5.2cm 1cm, clip=true, scale=0.44]{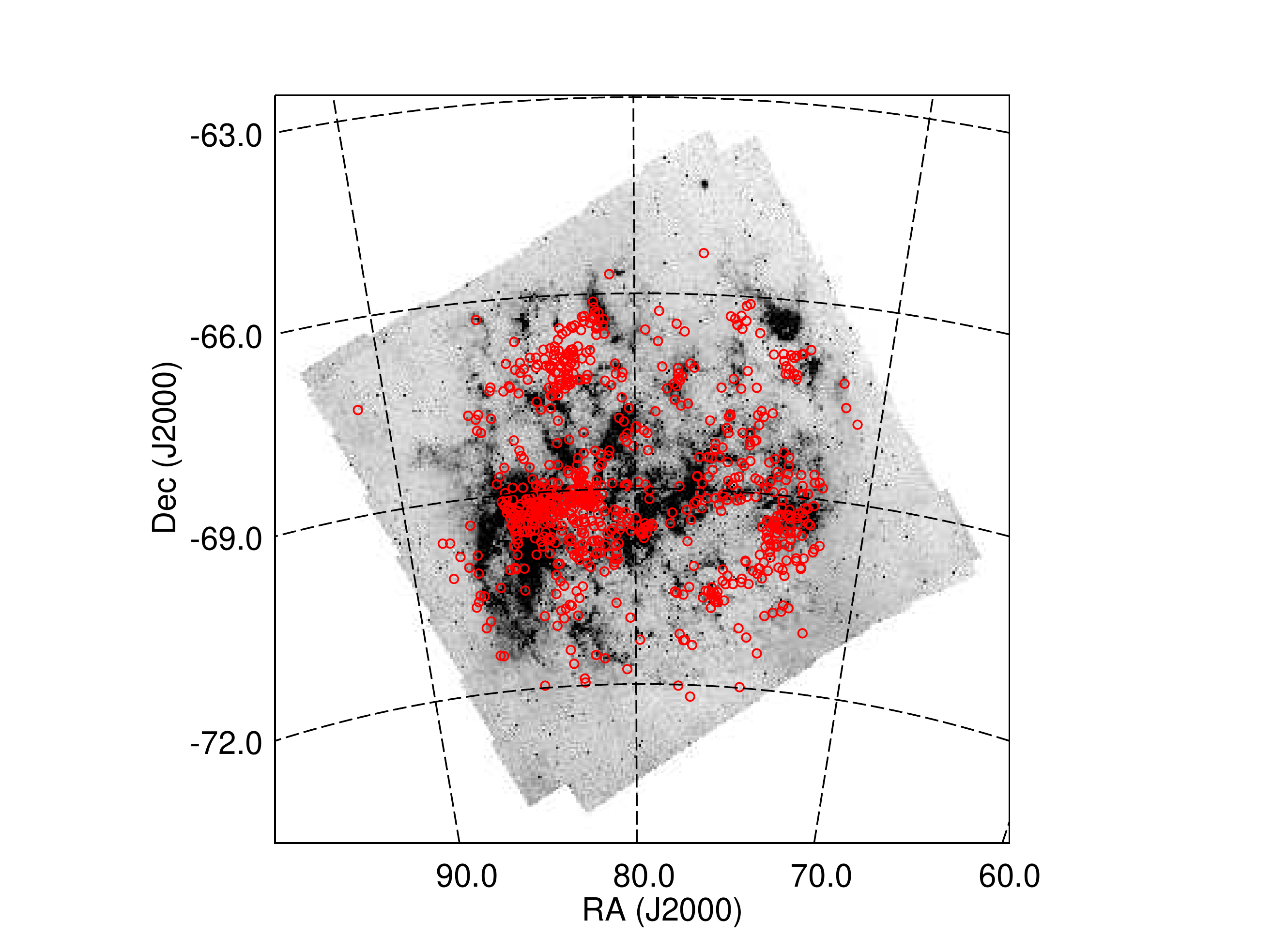}
\caption{Spatial distribution of 773 RSG candidates in the initial sample superposed on the Spitzer/SAGE 8 $\mu$m mosaic image. Most of the targets are close to the star formation region. \label{spatial}}
\end{figure}

Given all above, we have 189 targets from Paper \uppercase\expandafter{\romannumeral1}, 158 targets from M03, 505 targets from N12 (Category 1: LMC RSGs), 229 targets from G15 (heliocentric velocity $\geq$ 190~km/s, luminosity type Ia$\sim$Ib-II, spectral type late than G0), and 71 targets from J17. For simplicity, we have combined all the optical and NIR targets (from Paper \uppercase\expandafter{\romannumeral1}, M03, N12 and G15) together and removed the duplications by using a search radius of 3", which results in 749 unique sources. Then the MIR spectroscopic targets (from J17) have been compared with the optical and NIR targets to remove the MIR classified asymptotic giant branch stars (AGBs) without optical spectral counterparts (three targets) and to add new RSGs candidates (27 targets). Finally, there are 773 RSG candidates in our initial sample. The spatial distribution of all 773 RSG candidates is shown in Figure~\ref{spatial}.

During the compilation, several things draw our attention. First of all, there may be a confusion between RSGs and AGBs at faint magnitudes since they are both red and luminous, while RSGs are slightly brighter and less red. The separation of RSGs and AGBs by $M_{bol}=-7.1~mag$ has been proposed by \citet{Wood1983} but the most luminous AGBs, known as super-AGBs, can be as bright as $M_{bol}\sim-8.0~mag$ \citep{Herwig2005, Siess2006, Siess2007, Siess2010, Groenewegen2009, Doherty2016}. On the other hand, the heavily dust-obscured RSGs might be as faint and red as the average level of AGB stars. Secondly, not all of our targets are constrained by the systemic radial velocity of LMC, which means that a few of them might be foreground stars. Thirdly, there is a small discrepancy between optical and MIR spectroscopic classification. For example, one target ($R.A=83.72404$, $Decl.=-69.13391$) has been classified as a RSG in G15 but as an oxygen-rich AGB in J17. For such cases, the optical classifications have been chosen as the primary standards. In brief, a careful investigation is still needed to purify our sample as much as possible.

 \section{Photometry}

We collected the co-added infrared data of the initial sample by using a search radius of 3" from the Spitzer Enhanced Imaging Products (SEIP) source list, which contains sources detected with a high signal-to-noise ratio (S/N; ten-sigma level) in at least one channel among 12 NIR to MIR bands of J (1.25 $\mu$m), H (1.65 $\mu$m), $K_S$ (2.17 $\mu$m), IRAC1 (3.6 $\mu$m), IRAC2 (4.5 $\mu$m), IRAC3 (5.8 $\mu$m), IRAC4 (8.0 $\mu$m), MIPS24 (24 $\mu$m), WISE1 (3.4 $\mu$m), WISE2 (4.6 $\mu$m), WISE3 (12 $\mu$m) and WISE4 (22 $\mu$m), from 2MASS, Spitzer and Wide-field Infrared Survey Explorer (WISE; \citealt{Wright2010}). The $JHK_{S}$ data from 2MASS catalog is merged with the SEIP Source List using a match radius of 1", with the closest object being chosen, as well as the four bands of MIR data from ALLWISE catalog within a match radius of 3". It has been noticed that to ensure high reliability, strict cuts are placed on extracted sources in the SEIP source list. Some legitimate sources may appear to be missing due to cuts in size, compactness, blending, shape, and S/N, along with multiband detection requirements \footnote{http://irsa.ipac.caltech.edu/data/SPITZER/Enhanced/SEIP/docs/\\seip\_explanatory\_supplement\_v3.pdf}. In total, 772 targets have been selected from the SEIP source list with 773 entries. Only one target appears to have two very similar entries but without obvious blending from the image inspection. We removed the duplication. The $\sim$99.9\% single detection rate inside the search radius also implies that there would be no serious blending in our sample at these wavelengths. The one undetected target has been proven to be saturated in Spitzer bands and slightly offset from the given coordinate while it is still listed in the ALLWISE catalog. We retrieved the data for this target from the ALLWISE catalog instead of SEIP source list with the correct coordinate. All information about the initial sample is listed in Table~\ref{isample}. Targets without errors indicate a 95\% confidence upper limit. The wavelengths of 12 NIR to MIR bands, number of detected targets, and the related percentages in each band are listed in Table~\ref{tbl1}.

The MIR time-series data were collected from both ALLWISE and Near-Earth Object WISE Reactivation mission (NEOWISE-R; \citealt{Mainzer2014}). The AllWISE program combines data from the WISE cryogenic and NEOWISE \citep{Mainzer2011} post-cryogenic survey phases to generate new catalogs with enhanced photometric sensitivity and accuracy, and better astrometric precision compared to the 2012 WISE All-Sky Data Release. The NEOWISE-R is a three-year reactivation survey of WISE since 2013, which aims to study the population of near-Earth objects and comets in the 3.4 and 4.6 $\mu$m bands. Both ALLWISE and NEOWISE-R projects provide plenty of MIR time-series data which can be used in our study. However, since there are only two bands valid in NEOWISE-R, our study is focusing on the combination of ALLWISE and NEOWISE-R in WISE1 (3.4 $\mu$m) and WISE2 (4.6 $\mu$m) bands. A search radius of 3" has been applied to retrieve data from AllWISE Multiepoch Photometry Table and NEOWISE-R Single Exposure (L1b) Source Table, with following parameters used to filter out unreliable data points:
\begin{itemize}[noitemsep,topsep=0pt,parsep=0pt,partopsep=0pt]
  \item $qi\_fact>0$ -- Frameset image quality score, to exclude extractions from framesets that may have smeared images and subsequently biased photometry.
  \item $saa\_sep>0$ -- Angular distance from the nominal boundaries of the South Atlantic Anomaly (SAA), to avoid framesets taken when the WISE spacecraft was within the boundaries of the SAA. 
  \item $moon\_masked=0$ -- Moon masking flag, to avoid detections that are made on frames that may be contaminated by scattered moonlight.
  \item $qual\_frame>0$ -- Overall frameset quality score (NEOWISE-R only), to obtain detections from framesets with good quality.
  \item $det\_bit=3$ -- Detection flag (NEOWISE-R only), to obtain sources detected in both W1 and W2 bands.
\end{itemize} 

\begin{figure*}
\includegraphics[bb=115 370 430 690, scale=0.54]{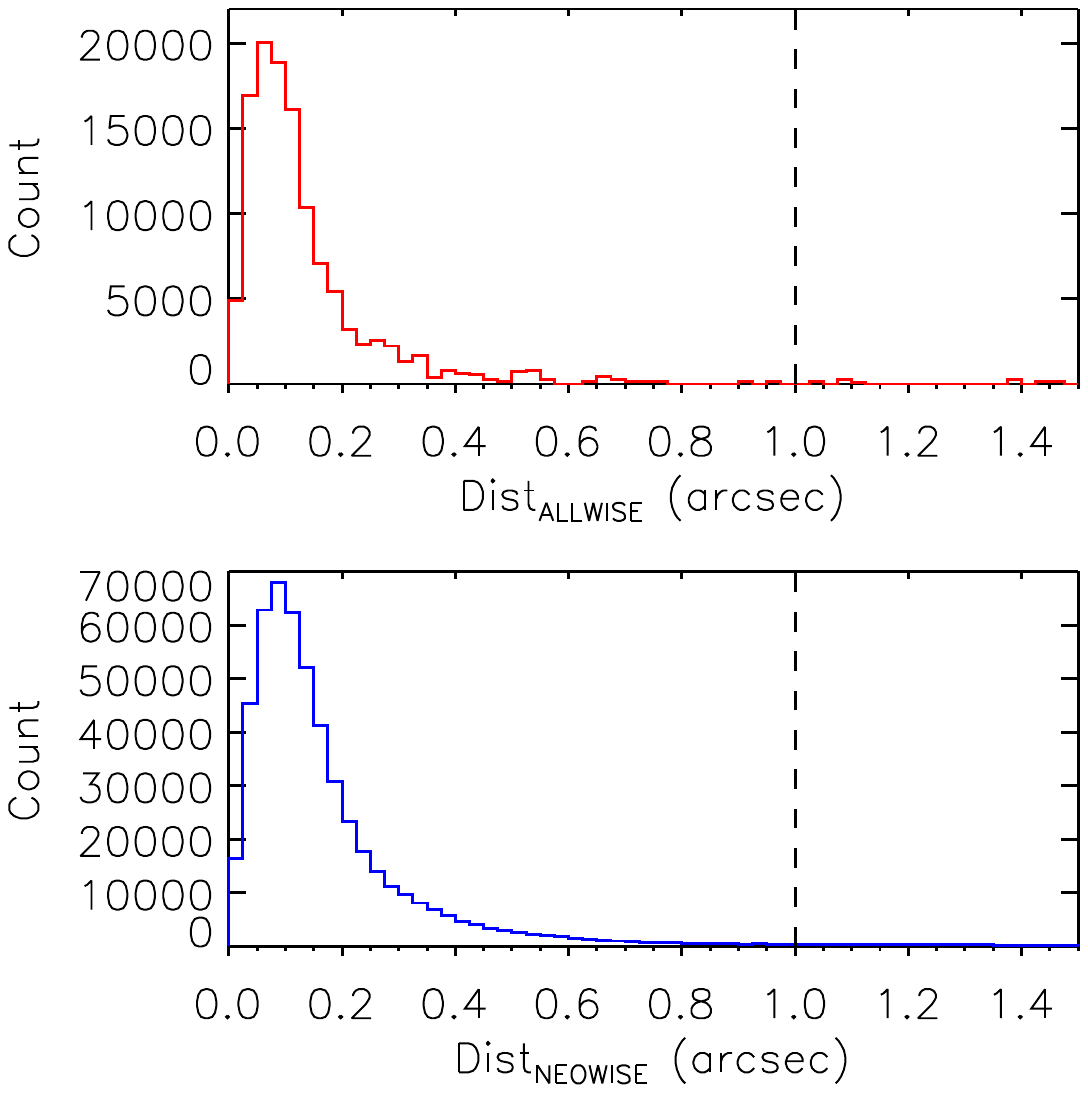}
\includegraphics[bb=115 370 430 690, scale=0.54]{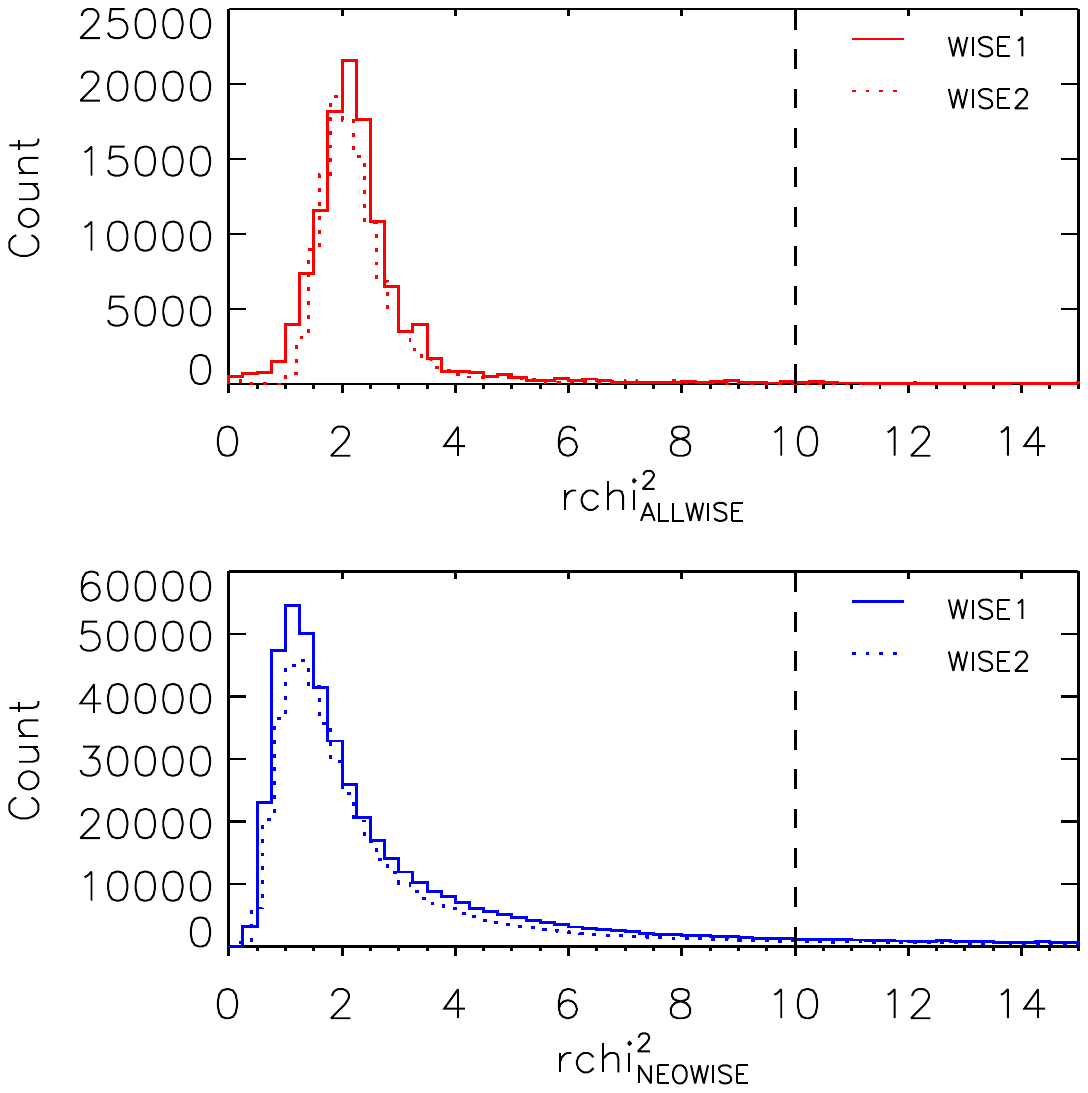}
\includegraphics[bb=115 370 430 690, scale=0.54]{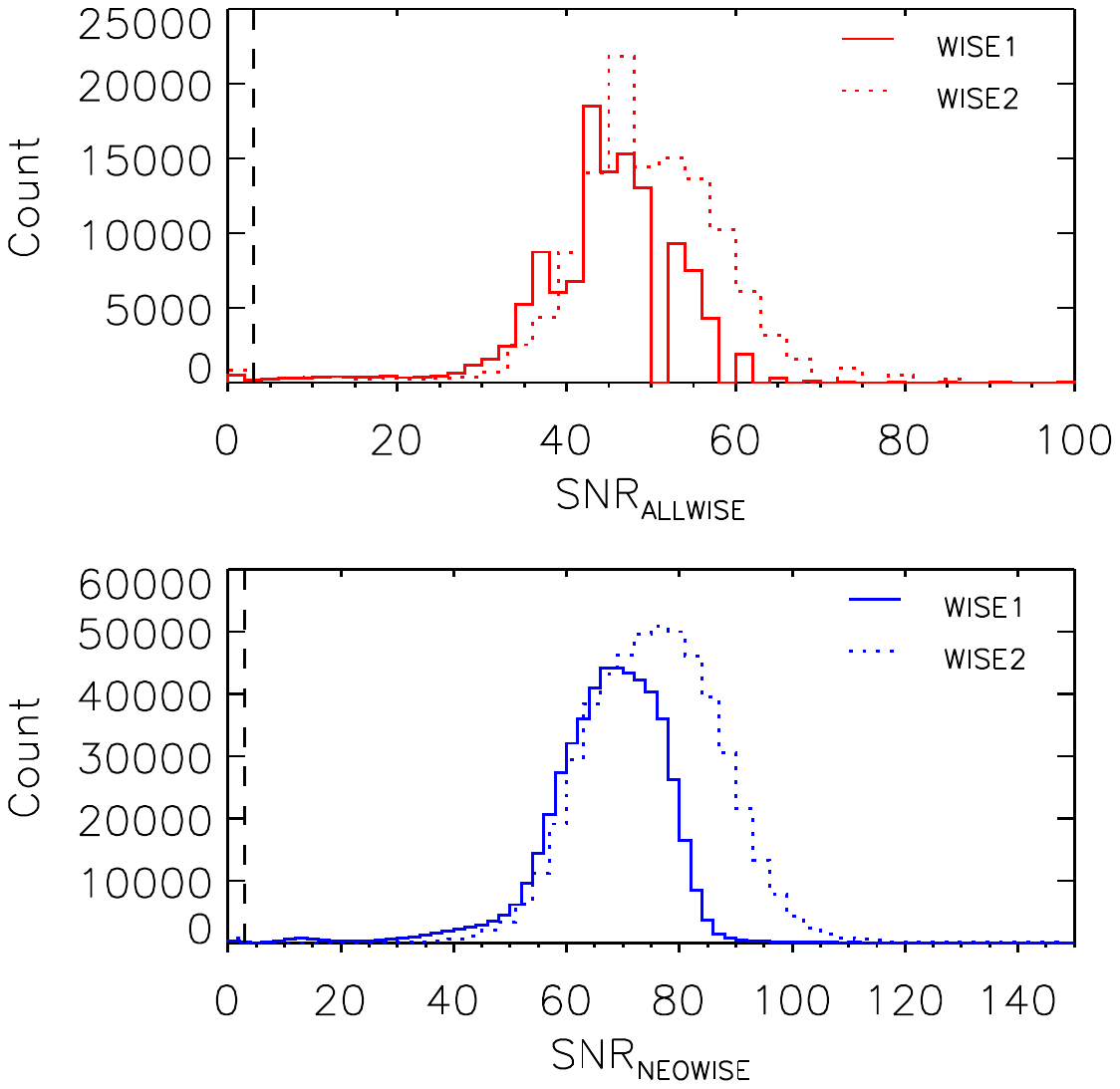}
\caption{Histograms of the positional offset parameter (left), the reduced $\chi^2$ of the profile-fit photometry (middle), and the S/N (right) for ALLWISE (top) and NEOWISE-R (bottom). The cut-offs are shown as dashed lines (1" for position, 10 for reduced $\chi^2$ and 3 for S/N). To distinguish between WISE1 and WISE2 bands, a slightly different bin size has been used for the WISE2 band. \label{stat}}
\end{figure*}

As well as the parameters used above, several cut-offs also have been applied to the positional offset parameter (to retrieve measurements within 1" from the given coordinates), reduced $\chi^2$ of the profile-fit photometry ($\chi^2\leq10$), and the S/N ($\geq3$) for each measurement as shown in Figure~\ref{stat}. 

\begin{figure}
\includegraphics[bb=85 410 460 635, scale=0.65]{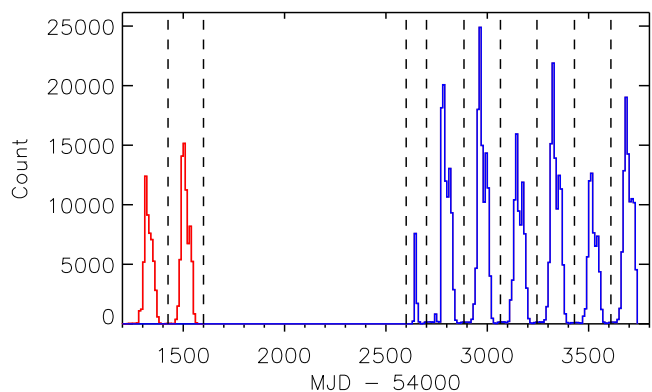}
\caption{Frame coverage of ALLWISE and NEOWISE-R. There are approximately nine major epochs with two epochs from ALLWISE (red) and seven epochs from NEOWISE-R (blue) separated by an approximately three-year gap. The dashed lines indicate the beginning and end of each epoch, as set by us. \label{epoch}}
\end{figure}

As WISE scans a fullsky area every half year, each part of the sky is revisited after about $\sim$180 days. Since WISE was placed in hibernation on February 2011 and reactivated on December 2013, there was also an approximately three-year gap in both WISE1- and WISE2-band data. Thus, the frame coverage is about nine major epochs spanning $\sim$2400 days ($\sim$6.6 years) with two epochs from ALLWISE and seven epochs from NEOWISE-R separated by the gap as shown in Figure~\ref{epoch}. The beginning and end of each epoch set by us are given in Table~\ref{tbl2}.

\begin{figure*}
\center
\includegraphics[bb=55 360 550 715, scale=0.9]{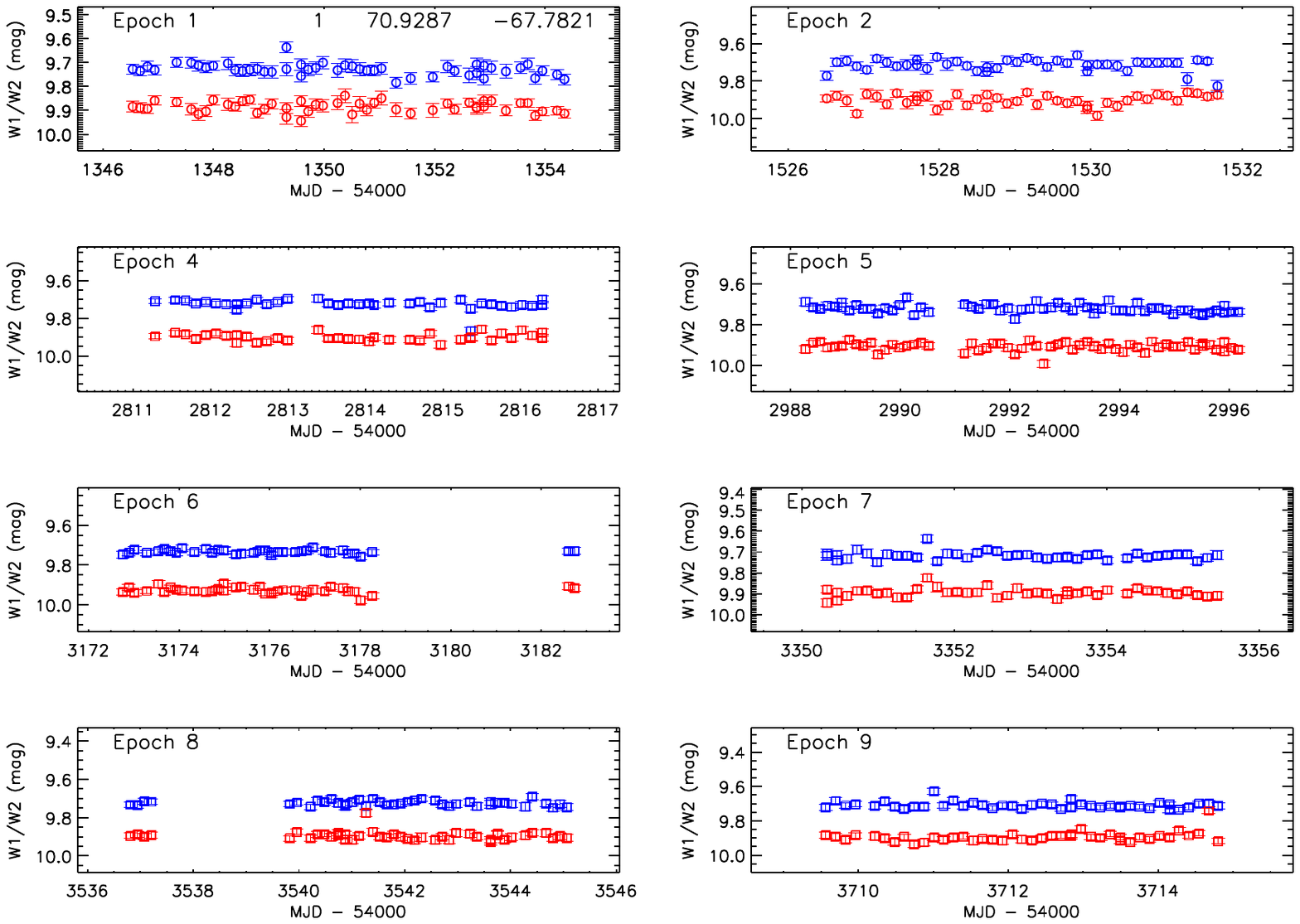}
\caption{Example of a MIR lightcuve in both WISE1 (blue) and WISE2 (red) bands. The ALLWISE and NEOWISE-R datasets are indicated by open circles and squares (same below), respectively. The ID, coordinate, and epoch are indicated on the diagram. \label{lc_single}}
\end{figure*}

\begin{figure*}
\includegraphics[bb=55 360 555 720]{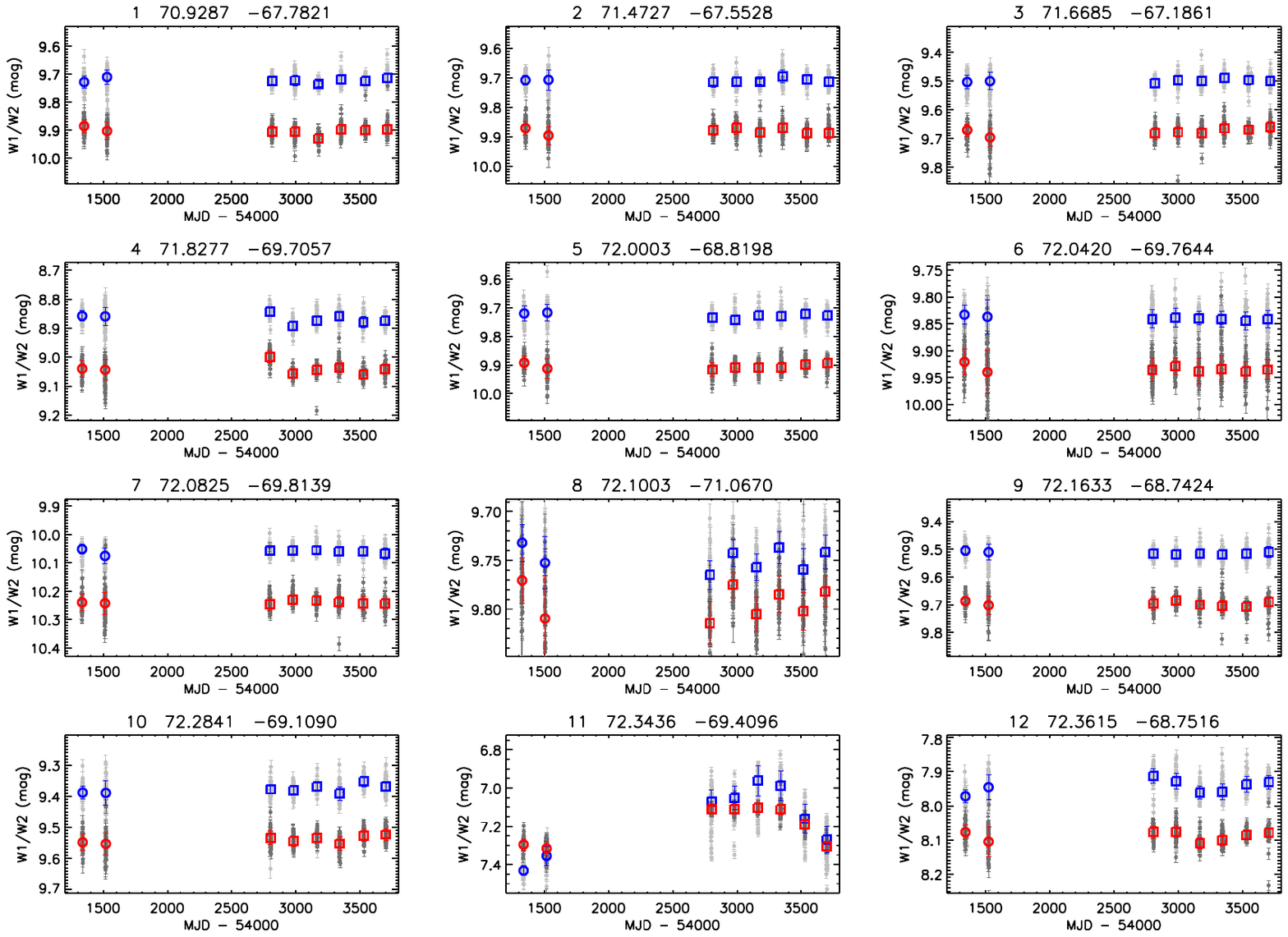}
\caption{Examples of original (light gray for WISE1 and dark gray for WISE2) and binned (blue for WISE1 and red for WISE2) lightcurves. The IDs and coordinates are indicated on the diagram. \label{lc_group}}
\end{figure*}

Due to the natural way of WISE allsky scanning, different parts of the sky are covered with different numbers of frames. For each target in our sample, the observation is actually conducted for approximately five to ten days during each epoch as shown in Figure~\ref{lc_single}. This time period is much less ($<5\%$) than the typical pulsation period of RSGs which is about 250 to 1000 days \citep{Kiss2006, Yang2011, Yang2012, Soraisam2018}. It is not even enough to derive a reliable period when combining all the epochs together. Thus, we have binned the data within each epoch by using the median values of the date and magnitudes. For each epoch, we require at least five valid points to calculate the median value. Figure~\ref{lc_group} shows the examples of the original lightcurves overlapped with the binned lightcurves. It can be seen that our procedure has almost no influence on the long-term variability of the targets.

\begin{figure}
\center
\includegraphics[bb=55 360 555 720, scale=0.5]{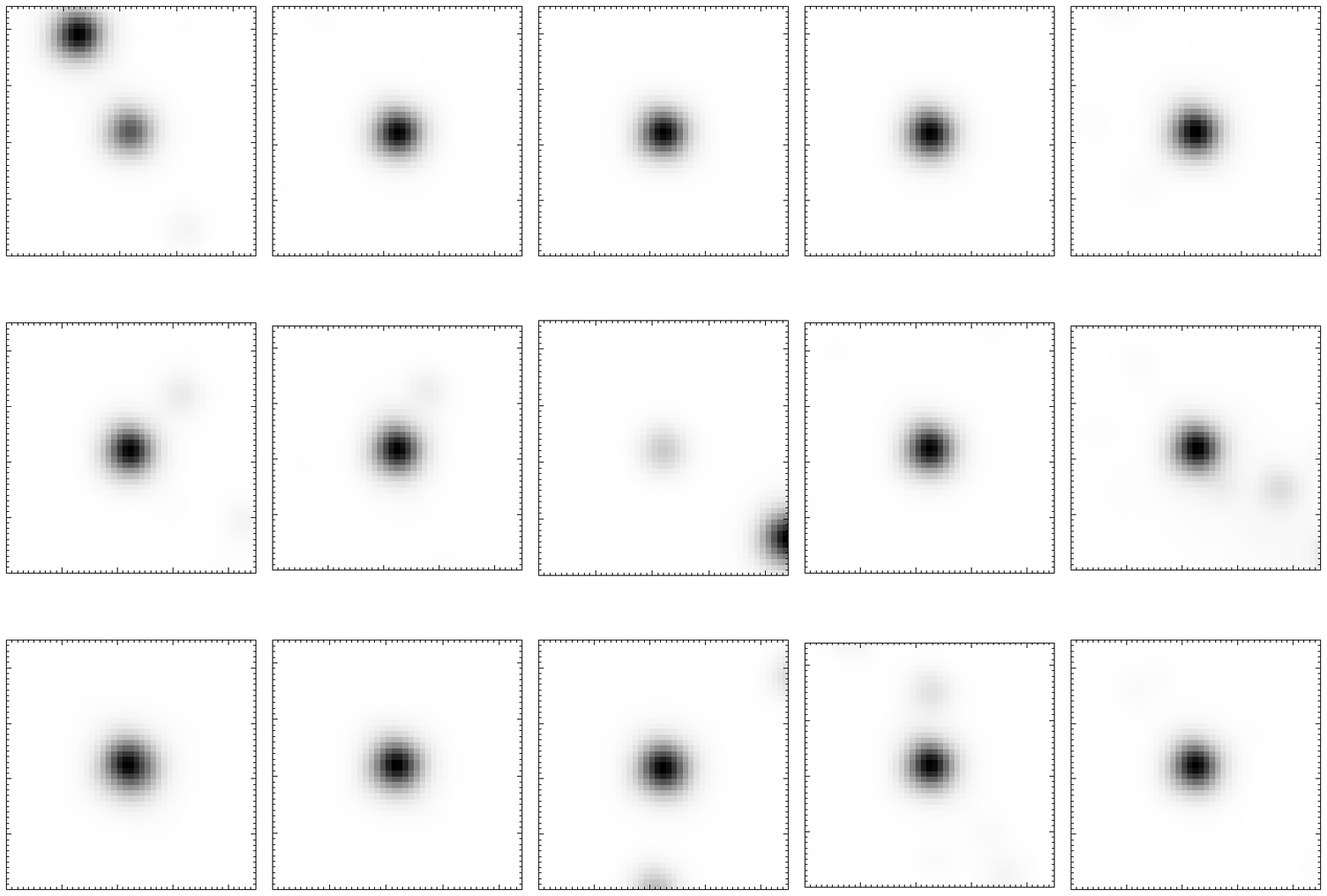}
\caption{Examples of WISE1-band images of the initial sample with a FOV of 60" for each image. \label{w1image}}
\end{figure}

Finally, we have visually inspected each target in WISE1-band image as shown in Figure~\ref{w1image}. The field-of-view (FOV) is 60", which indicates that the neighboring targets are actually far from each other and can be well separated by the moderate angular resolution ($\sim$6" in WISE1) of WISE.

\section{Data analysis}

As we stated above, the RSG sample could be contaminated by several factors. Thus, we attempted to address this problem with different approaches. 

\subsection{Color-magnitude diagrams}

We first used the color-magnitude diagram (CMD), which takes into account both brightness (magnitude) and effective temperature (color). We note that due to the large MLR of RSGs \citep{Chiosi1986, vanLoon1999, vanLoon2005, Mauron2011}, the position of RSGs on the diagram could be affected by high extinction from a dusty envelope or stellar wind. In order to avoid this effect as much as possible, the NIR and MIR bands have been used since the extinction in the infrared is much smaller than in the optical bands. 

\begin{figure}
\includegraphics[bb=125 365 505 690, scale=0.65]{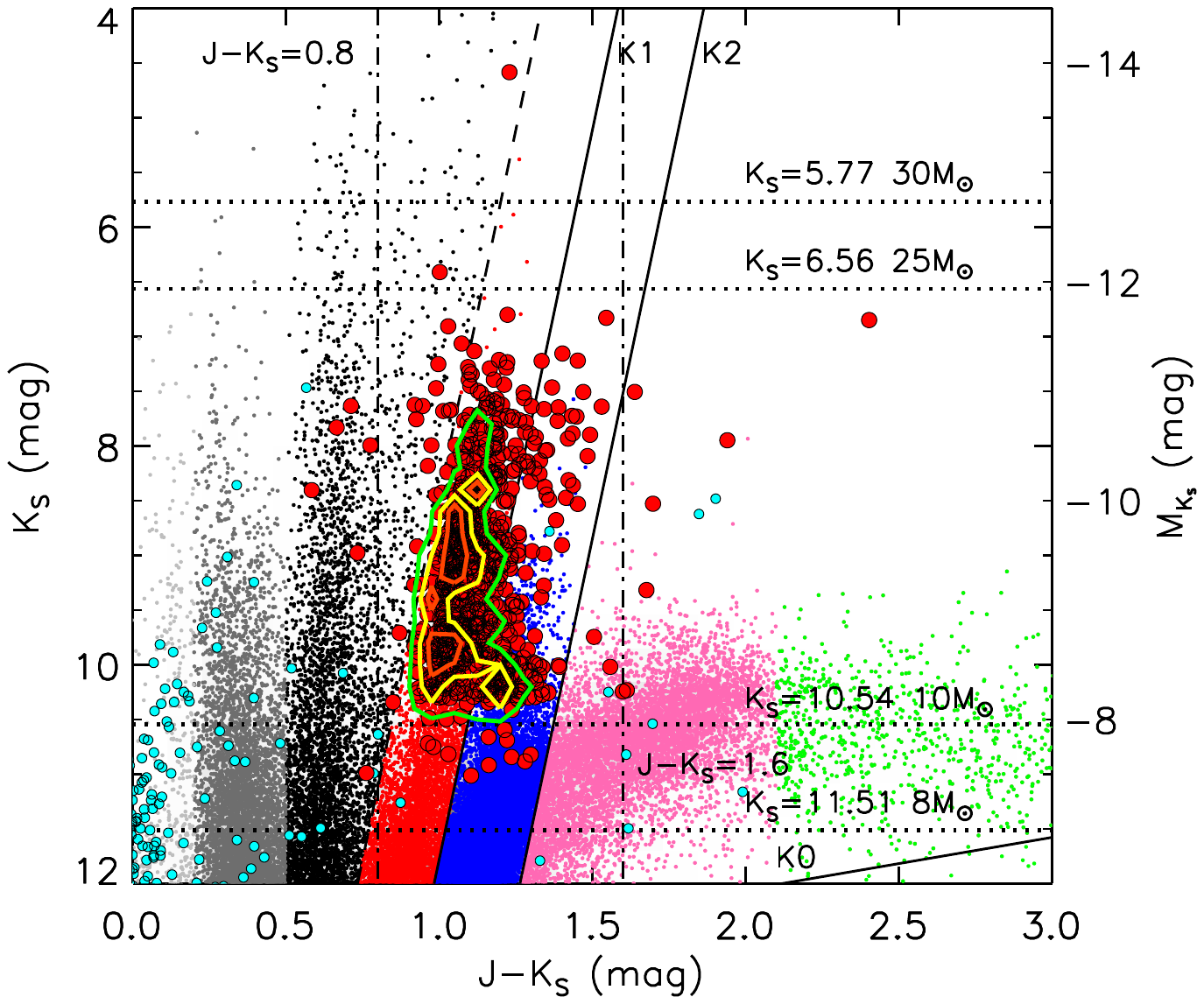}
\caption{$K_S$ versus $J-K_S$ diagram for the initial sample. For comparison, the C-AGBs (blue), O-AGBs (pink), x-AGBs (green) and RSGs (red) regions defined by the theoretical $J-K_S$ color cuts \citep{Cioni2006, Boyer2011}, and 1268 massive stars ($M\geq8~M_\sun$) in the LMC from \citet{Bonanos2009} (cyan) are also shown in the diagram. Four horizontal dotted lines indicate the theoretical luminosity limits of 8, 10, 25 and 30 $M_\sun$. The vertical dashed-dotted lines indicate the red boundary of carbon-rich stars as $J-K_S=1.6~mag$ suggested by \citet{Hughes1990} and a slightly strict blue boundary of $J-K_S=0.8~mag$. The 25\%, 50\% and 75\% density contour levels are given by the orange, yellow and green lines, respectively (same below). From the diagram, it can be seen that our sample agrees well with the luminosity limits and most targets are located in a region of $6.5<K_S<11.0~mag$ and $0.8<J-K_S<1.6~mag$, while about 71\% of RSGs candidates are inside the defined RSGs region. See text for details. \label{cmd_jhk}}
\end{figure}

Figure~\ref{cmd_jhk} shows $K_S$ versus $J-K_S$ diagram for the initial sample. The foreground extinction has not been taken into account since it is very small in J and $K_S$ bands ($A_J\approx$0.056 and $A_{K_S}\approx$0.022 mag; \citealt{Gao2013}), which is comparable to the observational uncertainty, if $E(B-V)\sim0.06~mag$ with the Galactic average value of $R_V=3.1$ is adopted \citep{Bessell1991, Oestreicher1995, Dobashi2008}. The 1268 massive stars ($M\geq8~M_\sun$) in the LMC from \citet{Bonanos2009} (cyan), which are collected from the literature, have been added for comparison. Background targets are selected by $70\leq R.A.\leq 90$ deg, $-72.5\leq Decl. \leq-64.5$ deg and $K_S\leq12.5~mag$ which cover almost the whole LMC.

Following Paper \uppercase\expandafter{\romannumeral1}, we set up the theoretical limits of luminosities for our RSGs candidates as $8^4-30^4~L_\sun$, which correspond to $8-30~M_\sun$ converted by $L/L_\sun=(M/M_\sun)^\gamma$ with $\gamma$ very close to 4.0 \citep{Stothers1971}. It then can be converted to the absolute bolometric magnitude ($M_{bol}$) in the range of -4.29 to -10.03 mag by $M_{bol}=4.74-2.5log(L/L_\sun)$. The $M_{bol}$ is converted to the $K_S$-band magnitude of 11.51 to 5.77 mag by using a constant $K_S$-band bolometric correction ($BC_{K_S}$) of 2.69 from \citet{Davies2013} with a distance modulus of 18.493 for the LMC \citep{Pietrzynski2013}. In addition, the limits of 10 ($K_S=10.54~mag$) and 25 $M_\sun$ ($K_S=6.56~mag$) also have been shown in the diagram for comparison. It has been noted that our sample is in good agreement with the luminosity limits and almost all targets are located within the range of $8-30~M_\sun$, in which most of the targets are between 10 and 25 $M_\sun$ with a few fainter or brighter ones. Only one outlier (No.773) is far brighter than the upper limit of 30 $M_\sun$, which may not be a RSGs. This also solves a long puzzle in the Paper \uppercase\expandafter{\romannumeral1} that a previous huge gap between observed sample and theoretical predication now has been fulfilled (see Figure 2 of Paper \uppercase\expandafter{\romannumeral1}). We also expect that there will be more low-luminosity RSGs in the range of 11.0 to 10.5 mag but higher contamination from galactic dwarfs may also appear.

Besides the luminosity limits, the theoretical $J-K_S$ color cuts from \citet{Cioni2006} are shown in the diagram, with C-rich AGBs (C-AGBs; blue) defined by $K2<K_S<K0$, O-rich AGBs (O-AGBs; pink) defined by $K_S<K0$ and $K1<K_S<K2$, extreme-AGBs (x-AGBs; green) defined by $K_S<K_S$-band tip of the red giant branch ($K_S-TRGB\approx12~mag$) and $J-K_S>2.1~mag$. The definition of the RSGs region follows \citet{Boyer2011} by $\Delta(J-K_S)=0.25~mag$ from the O-AGBs shown as the dashed line in the diagram. The vertical dashed-dotted lines indicate the red boundary of carbon-rich stars as $J-K_S=1.6~mag$ suggested by \citet{Hughes1990} and a slightly strict blue boundary of $J-K_S=0.8~mag$. Moreover, the foreground region (black), foreground plus A-G supergiants region (dark gray) and the OB stars region (gray) are also shown in the diagram \citep{Nikolaev2000, Boyer2011}. From the diagram, most of our targets are located in a region of $6.5<K_S<11.0~mag$ and $0.8<J-K_S<1.6~mag$, while about 71\% of RSGs candidates are inside the defined RSGs region with an extension to the O-AGBs region which may be due to reddening, the intrinsic lower effective temperature or the photometric accuracy.

Following the above, we also show a few examples of the deepest and brightest ($K_S$/WISE1/IRAC1) band-based multiple CMDs (for the bluest and reddest colors) in Figure~\ref{cmd_sample} (the whole set of CMDs can be found in the Appendix as $K_S$ versus $K_S-WISE1$, $K_S-IRAC1$, $K_S-IRAC2$, $K_S-WISE2$, $K_S-IRAC3$, $K_S-IRAC4$, $K_S-WISE3$, $K_S-WISE4$ and $K_S-MIPS24$ in Figure~\ref{ks_cmd}, WISE1 versus $J-WISE1$, $H-WISE1$, $K_S-WISE1$, $WISE1-IRAC1$, $WISE1-IRAC2$, $WISE1-WISE2$, $WISE1-IRAC3$, $WISE1-IRAC4$, $WISE1-WISE3$, $WISE1-WISE4$ and $WISE1-MIPS24$ in Figure~\ref{w1_cmd}, and IRAC1 versus $J-IRAC1$, $H-IRAC1$, $K_S-IRAC1$, $WISE1-IRAC1$, $IRAC1-IRAC2$, $IRAC1-WISE2$, $IRAC1-IRAC3$, $IRAC1-IRAC4$, $IRAC1-WISE3$, $IRAC1-WISE4$ and $IRAC1-MIPS24$ in Figure~\ref{i1_cmd}). Due to the multiplicity of passbands and the complexity of filter combinations, it is difficult to identify outliers, which may or may not be RSGs, in every filter combination by eye. Therefore, we adopt the method used in the Hubble Catalog of Variables (Bonanos et al., in preparation; \citealt{Gavras2017, Yang2017}) to identify outliers. In each diagram, the targets have been divided into ten overlapped magnitude bins with at least 5\% of targets included in each bin to maintain the statistical significance. The mean ($C_{mean}$) and standard deviation ($C_{\sigma}$) of each color index have been calculated within each bin. Targets with absolute value of color index larger than $C_{mean}+3C_{\sigma}$ in two consecutive bins (except the first and last bins) or too bright ($K_S/WISE1/IRAC1<5.5~mag$)/too faint ($K_S/WISE1/IRAC1>11.5~mag$) are defined as outliers. Still, since the sensitivities, effective wavelengths, bandwidths, photometric accuracies and completenesses between these infrared surveys are different and the physical mechanism behind the evolution of RSGs are unclear (e.g., interior structure, MLR and variability), some targets may appear to be outliers in one CMD but not in others. Thus, we define here a CMD outlier should be identified in more than half of $K_S/WISE1/IRAC1$-based multiple CMDs. It has to be emphasized that this is an arbitrary decision rather than a statistical prediction considering the complex datasets we have. The names of CMDs and IDs of outliers in $K_S/WISE1/IRAC1$-based CMDs are listed in Tables~\ref{ks_outlier},~\ref{w1_outlier}, and~\ref{i1_outlier}, respectively. In total, there are 11 CMD outliers of No.22, 77, 204, 283, 291, 376, 394, 592, 679, 723, 773. Another thing that caught our attention is that, at the longer wavelengths, the color indexes of the foreground population seem to be around zero. Targets Nos.123 and 547 are classified as foreground outliers shown as dark yellow open squares, due to their constant appearance in the foreground population region, especially in the longer wavelengths. From the diagrams, it also seems that there are two populations (brighter and fainter populations, approximately separated by the 25\% contour level) of RSGs with different distributions and tendencies which will be discussed in the later sections.

\begin{figure*}
\center
\includegraphics[bb=55 365 560 705]{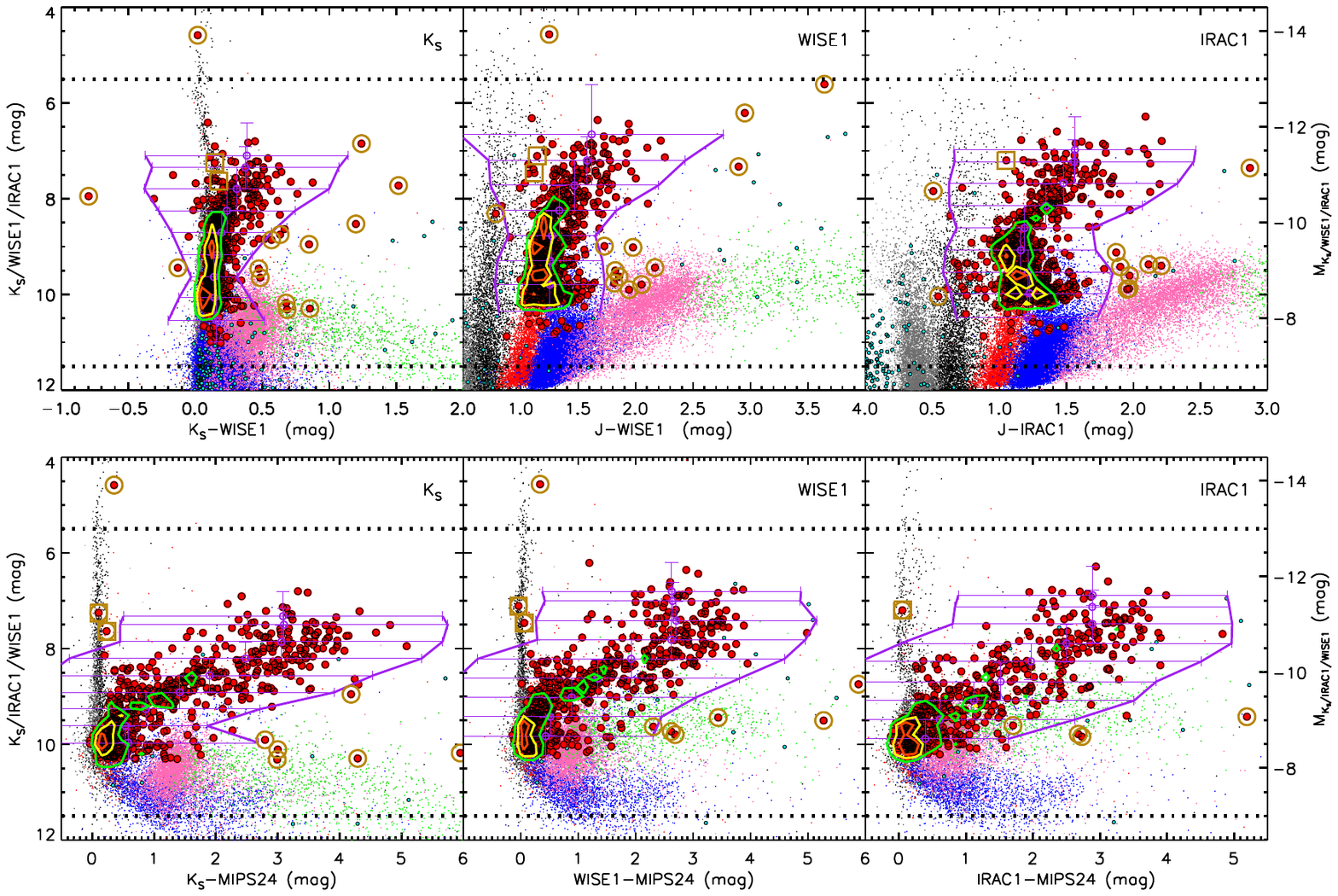}
\caption{Same as Fig~\ref{cmd_jhk} but for $K_S$ versus $K_S-WISE1/K_S-MIPS24$, $WISE1$ versus $J-WISE1/WISE1-MIPS24$, $IRAC1$ versus $J-IRAC1/IRAC1-MIPS24$. The luminosity and color limits are shown as black dotted lines and purple solid lines in each diagram. Outliers (dark yellow open circles) are defined by absolute value of color indexes larger than $C_{mean}+3C_{\sigma}$ in two consecutive bins or too bright or too faint (see text for details). Targets with 95\% confidence upper limit are shown as red open circles (same below). \label{cmd_sample}}
\end{figure*}

\subsection{Color-color diagrams and the reddening-free Q parameters}

\begin{figure*}
\center
\includegraphics[bb=80 360 495 715]{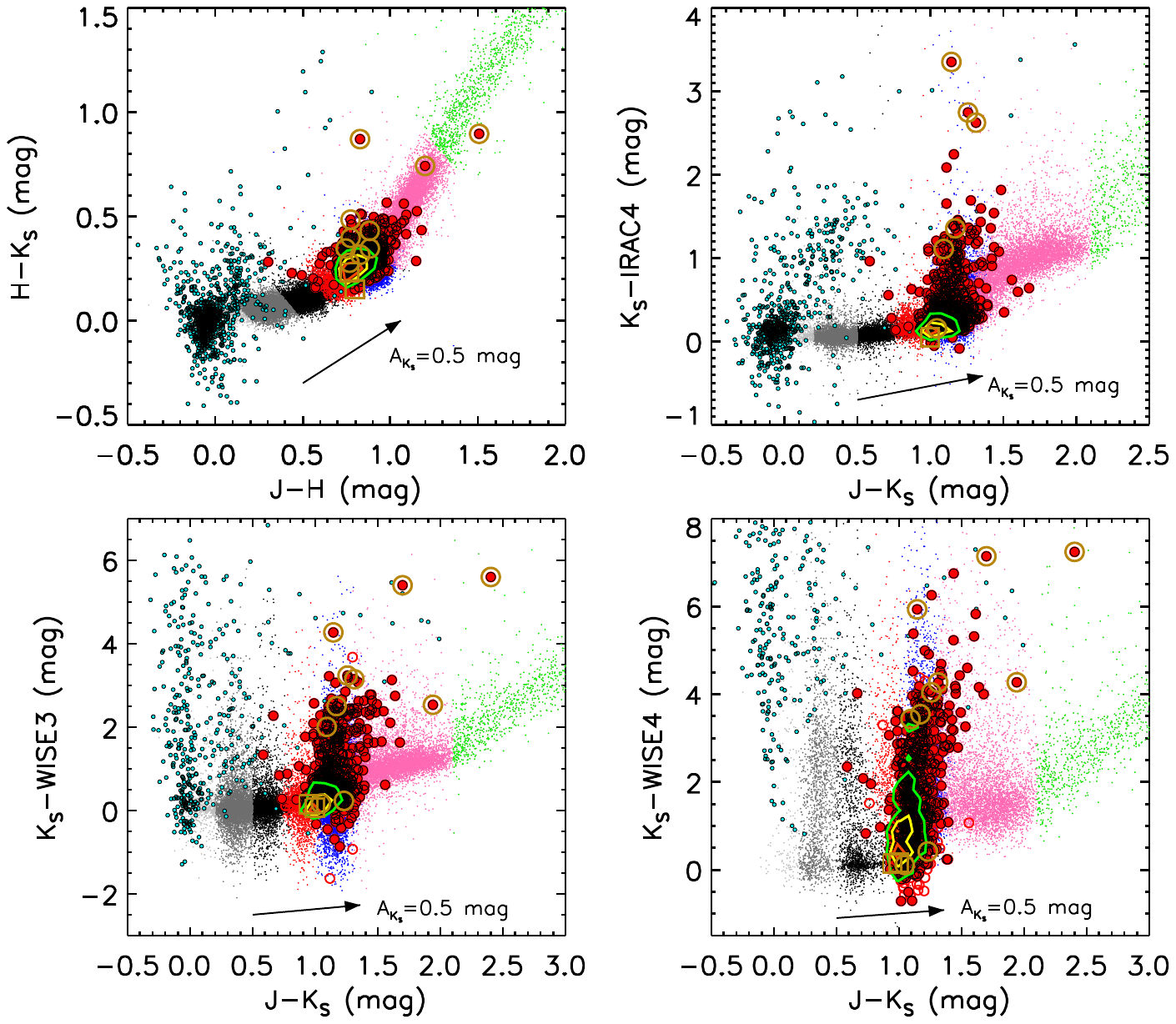}
\caption{CCDs of $J-H$/$H-K_S$, $J-K_S$/$K_S-IRAC4$, $J-K_S$/$K_S-WISE3$ and $J-K_S$/$K_S-WISE4$ for the initial sample. An $A_{K_S}\sim0.5~mag$ reddening vector is shown in the diagram for reference (same below). The two populations of RSGs (brighter and fainter) seen in previous CMDs also appearing in $J-K_S$/$K_S-IRAC4$ and $J-K_S$/$K_S-WISE3$ diagrams due to the higher sensitivity in IRAC4/WISE3 compared to WISE4 band, with fainter population is clumped around the infrared excess turning point and the brighter population is extended toward the red end of $K_S-IRAC4$/$K_S-WISE3$ color. \label{ccd_q}}
\end{figure*}

\begin{figure*}
\center
\includegraphics[bb=90 360 490 715]{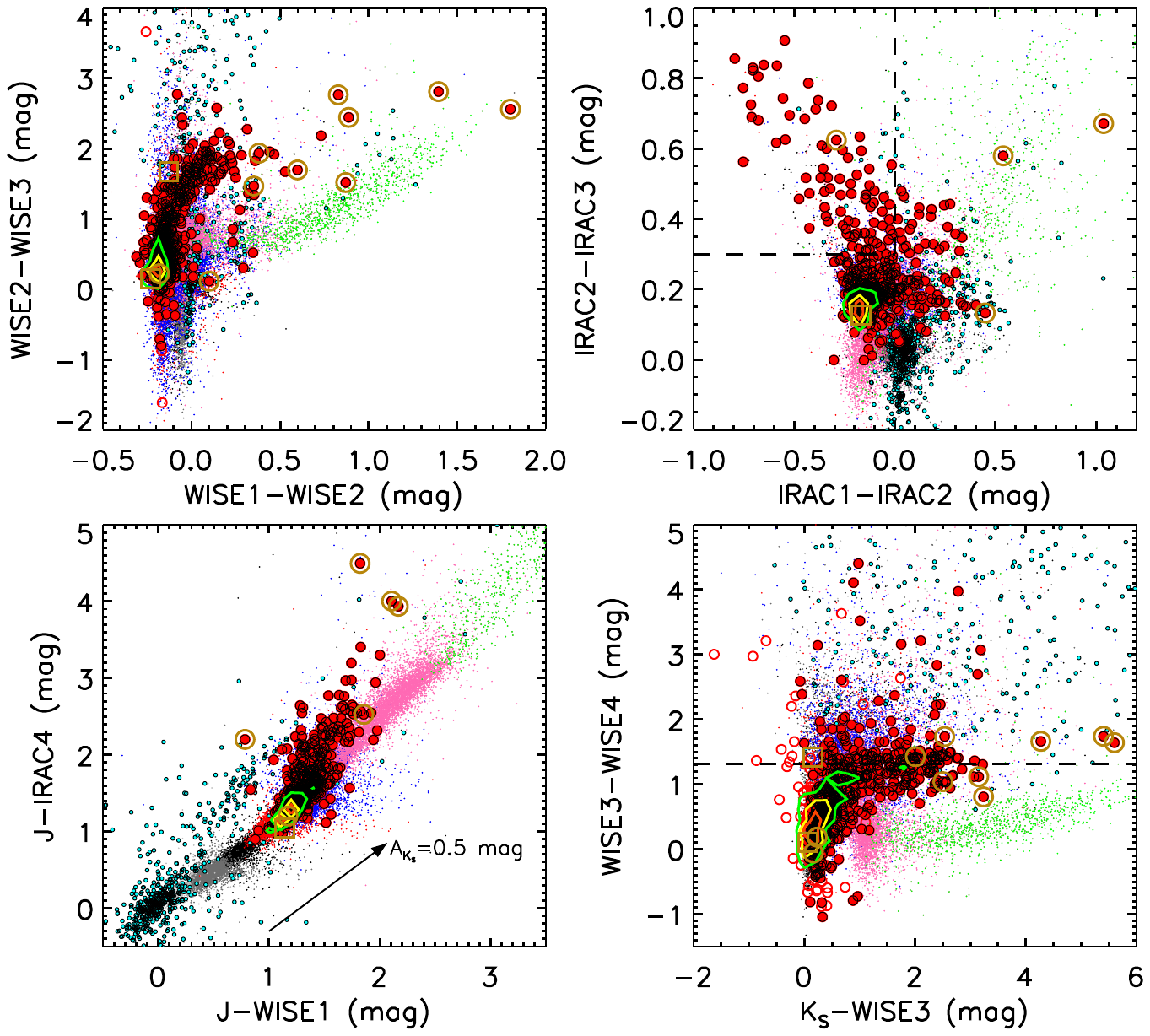}
\caption{Same as Fig~\ref{ccd_q} but for $WISE1-WISE2$/$WISE2-WISE3$, $IRAC1-IRAC2$/$IRAC2-IRAC3$, $J-WISE1$/$J-IRAC4$ and $K_S-WISE3$/$WISE3-WISE4$. A group of targets ($\sim15\%$, upper left region covered by the dashed lines) with probable PAH feature show redder color in $IRAC2-IRAC3$ and even bluer color in $IRAC1-IRAC2$ than the others in the $IRAC1-IRAC2$/$IRAC2-IRAC3$ diagram. The dashed line ($WISE3-WISE4\approx1.3~mag$) in the $K_S-WISE3$/$WISE3-WISE4$ diagram indicates the color where $F_\mu(WISE4)=F_\mu(WISE3)$. \label{ccd_other}}
\end{figure*}

Except for the CMDs, the color-color diagrams (CCDs) also have been investigated as shown in Figures~\ref{ccd_q} and \ref{ccd_other} with the CMD and foreground outliers indicated as before. Figure~\ref{ccd_q} shows the NIR and MIR color planes related to the $K_S$-band as $J-H$/$H-K_S$, $J-K_S$/$K_S-IRAC4$, $J-K_S$/$K_S-WISE3$ and $J-K_S$/$K_S-WISE4$. The reddening vector is adopted by using a precise measure of the MIR interstellar extinction law from \citet{Xue2016}, who select a large sample of Galactic G- and K-type giants ($3600\leq T_{eff}\leq 5200$ K with a step of 100 K, which is similar to the intrinsic temperature of RSGs), based on the stellar parameters derived from the SDSS-III/Apache Point Observatory Galaxy Evolution Experiment (APOGEE) spectroscopic survey \citep{Eisenstein2011}, to determine the MIR extinction for WISE, Spitzer/IRAC, Spitzer/MIPS24 and AKARI/S9W bands (see \citealt{Xue2016} for details). Here we have to emphasize that this is a very good approximation but may not be fully accurate, due to the difference in the metallicity between the LMC and our Milky Way. Since previous extinction maps show that the maximum $A_V$ of the LMC is close to $\sim5~mag$ \citep{Zaritsky2004, Imara2007, Dobashi2008}, considering $A_{K_S}/A_V\sim0.1$, we use $A_{K_S}\sim0.5~mag$ as a reference shown in the diagram \citep{Gao2013, Zhao2018}. For convenience, the reddening vector is not shown in some diagrams since it is too small to see. Here the interstellar reddening will move stars along the direction of the reddening vector, while circumstellar reddening makes stars deviate from it \citep{Bonanos2010, Wachter2010, Messineo2012}. It can be seen from the diagrams that the RSG population is located on the turning point of the infrared excess, where the reddening changes from interstellar dominated to the circumstellar dominated, and shows different tendency than the C-AGBs and X-AGBs, which may be mainly caused by the chemical composition and/or different MLRs. Due to the similar intrinsic temperatures and chemical composition, it is also overlapped with the O-AGBs which makes it indistinguishable in these diagrams. In the NIR plane ($J-H$/$H-K_S$), the majority of the RSGs candidates are clumped at the range of $0.6<J-H<1.1~mag$ and $0.1<H-K_S<0.5~mag$, which is consistent with the result of \citet{Rayner2009}. At longer wavelengths ($J-K_S$/$K_S-IRAC4$, $J-K_S$/$K_S-WISE3$ and $J-K_S$/$K_S-WISE4$), the increasing circumstellar reddening causes large excesses of about 2.0, 4.0 and 6.0 mag in $K_S-IRAC4$, $K_S-WISE3$ and $K_S-WISE4$, respectively. In particular, the two populations (brighter and fainter) of RSGs seen in previous CMDs also appearing in $J-K_S$/$K_S-IRAC4$ and $J-K_S$/$K_S-WISE3$ diagrams due to the higher sensitivity in IRAC4/WISE3 compared to WISE4 band, with fainter population is clumped around the infrared excess turning point and the brighter population is extended toward the red end of $K_S-IRAC4$/$K_S-WISE3$ color.

Several other CCDs of $WISE1-WISE2$/$WISE2-WISE3$, $IRAC1-IRAC2$/$IRAC2-IRAC3$, $J-WISE1$/$J-IRAC4$ and $K_S-WISE3$/$WISE3-WISE4$ have been shown in the Figure~\ref{ccd_other}. For the upper two diagrams considering only WISE or IRAC colors, the blue color of RSGs population in $WISE1-WISE2$/$IRAC1-IRAC2$ is mainly caused by the CO absorption around 4.6 $\mu$m~that can severely depress the flux for the sources with less infrared excess \citep{Verhoelst2009, Boyer2011, Reiter2015}. With the increasing infrared excess, targets are gradually turning into redder color as can be clearly seen in the $WISE1-WISE2$/$WISE2-WISE3$ diagrams, where two populations (brighter and fainter) of RSGs appear to be separated by $WISE2-WISE3\approx1.1~mag$. Interestingly, while the majority of targets clump around $IRAC1-IRAC2\approx-0.1~mag$ and $IRAC2-IRAC3\approx0.2~mag$ in the $IRAC1-IRAC2$/$IRAC2-IRAC3$ diagram, there is another group of targets ($\sim15\%$, upper left region covered by the dashed lines) showing redder color in $IRAC2-IRAC3$ and even bluer color in $IRAC1-IRAC2$ than the others. The redder color of $IRAC2-IRAC3$ is likely due to the polycyclic aromatic hydrocarbon (PAH) emission in 6.3 $\mu$m~partially captured by the IRAC3 filter \citep{Buchanan2006} and in agreement with the percentage found by \citet{Verhoelst2009}. This is also further confirmed by the $IRAC2-IRAC3$/$IRAC3-IRAC4$ diagram (not shown here), in which IRAC4 traces the PAH emission in 7.7 and 8.6 $\mu$m. In addition, the redder color of $IRAC2-IRAC3$ is also correlated with the extreme blue color of $IRAC1-IRAC2$, which could be explained by the PAH emission in 3.3 $\mu$m covered by the IRAC1 band. In brief, the population of O-rich RSGs showing PAH emission may need further investigation. Still, except for the RSGs with PAH features, there are no significant differences between the colors of O-AGBs and RSGs since they are tracing the same color sequence. For the $J-WISE1$/$J-IRAC4$ diagram, the $J-WISE1$ color can be considered as a good indicator of spectral type \citep{Gonzalez2015}, while the $J-IRAC4$ color represents the infrared excess related to the 10 $\mu$m~silicate emission feature, which is attributed to stretching of the Si-O bonds, partially captured by the IRAC4 band \citep{Sloan2008, Bonanos2009, Boyer2011}. There are $\sim1.0~mag$ and $\sim2.0~mag$ spread in $J-WISE1$ and $J-IRAC4$ color, respectively. Both colors are getting redder which indicates that the infrared excess is increasing with the spectral type. For the last diagram, the $K_S-WISE3$ color is similar to the MLR indicator of $K-[12]$ \citep{Josselin2000}, while the $WISE3-WISE4$ indicates the MIR excess. The dashed line ($WISE3-WISE4\approx1.3~mag$) shows the limit of $F_\mu(WISE4)=F_\mu(WISE3)$. As can be seen from the diagram, there is a flat MIR tail with the increasing MLR, which is likely due to the combination of silicate emission in 9.7 and 18 $\mu$m~and the dust emission at the longer wavelengths \citep{Buchanan2006}. 

\citet{Messineo2012} used the Q1 and Q2 parameters, which were the measurements of the deviation from the reddening vector in the $J-H$ versus $H-K_S$ and $J-K_S$ versus $K_s-[8.0]$ planes, to separate RSGs, W-R stars, luminous blue variables (LBVs) from AGBs. Thanks to the precise measurement of MIR interstellar extinction law from \citet{Xue2016}, we could now also be able to construct arbitrary dereddening parameters related to the IRAC4, S9W, WISE3, WISE4, and MIPS24 bands. The advantage of this method is the full sky coverage of WISE and AKARI compared to Spitzer. Still, we need to emphasize that the metallicity and photometric accuracy might be the key issues when the method is applied to different fields. For our case, due to the strict cut-offs in SEIP source list, only $\sim$90\% of targets are valid in IRAC4 (8 $\mu$m), while $\sim$99\% of targets are valid in WISE3 (12 $\mu$m). Based on Figure~\ref{ccd_q}, here we define new parameters called Q3 and Q4 as,
\begin{equation}
Q3=(J-K_S)-3.717*(K_S-WISE3),
\end{equation}
\begin{equation}
Q4=(J-K_S)-2.703*(K_S-WISE4).
\end{equation}
As stated by \citet{Xue2016}, the result on the WISE4 band must be treated cautiously, since the statistical uncertainty (determined by the bootstrap resampling and Monte-Carlo simulation) of WISE4 is one order of magnitude bigger than the other bands, which may be due to the relatively low sensitivity and the small number of sample crossmatched with the APOGEE survey. Thus, the Q4 parameter defined here is only used as a reference. 

\begin{figure*}
\center
\includegraphics[bb=95 360 485 715]{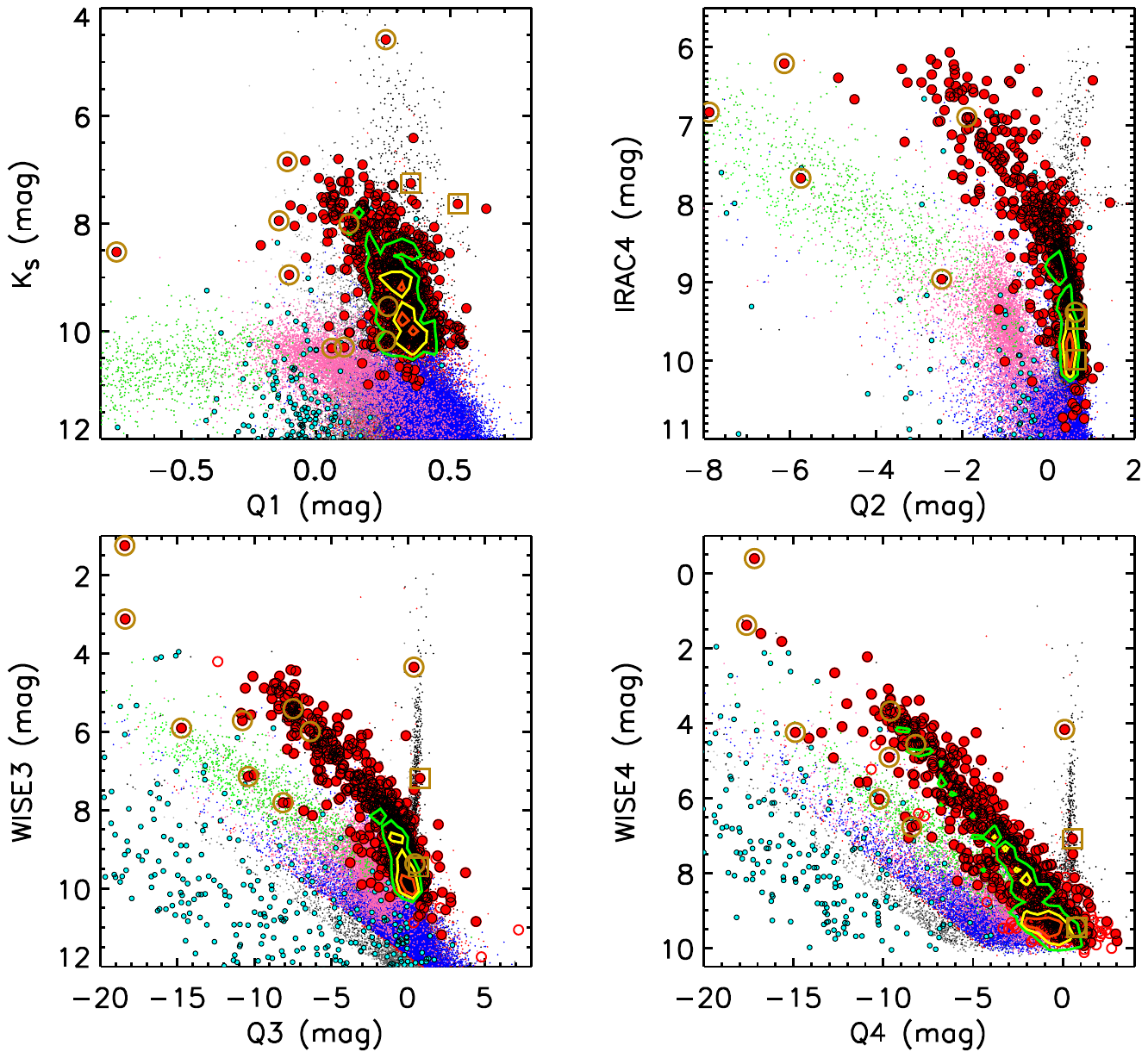}
\caption{Four Q parameters of the initial sample, with Q1 and Q2 defined by \citet{Messineo2012}, Q3 and Q4 defined by us. The RSG population appears to be in a separate branch between foreground dwarfs and AGBs in the bright end. Due to the relatively low sensitivity and the small number of sample crossmatched with the APOGEE survey in \citet{Xue2016}, the Q4 parameter defined here is only used as a reference. \label{cmd_q}}
\end{figure*}

Figure~\ref{cmd_q} shows all four Q parameters of the initial sample, with Q1 and Q2 defined by \citet{Messineo2012}, Q3 and Q4 defined by us. It can be seen that, due to the increasing MLR, at the longer wavelength, the RSG population appears to be in a separate branch between foreground stars and AGBs. We also note that, a more clear separation only appears for the brighter RSGs with larger circumstellar reddening. The faint end is still mixed with other populations. 

\subsection{Spectral energy distribution}

\begin{figure}
\center
\includegraphics[bb=185 400 395 675]{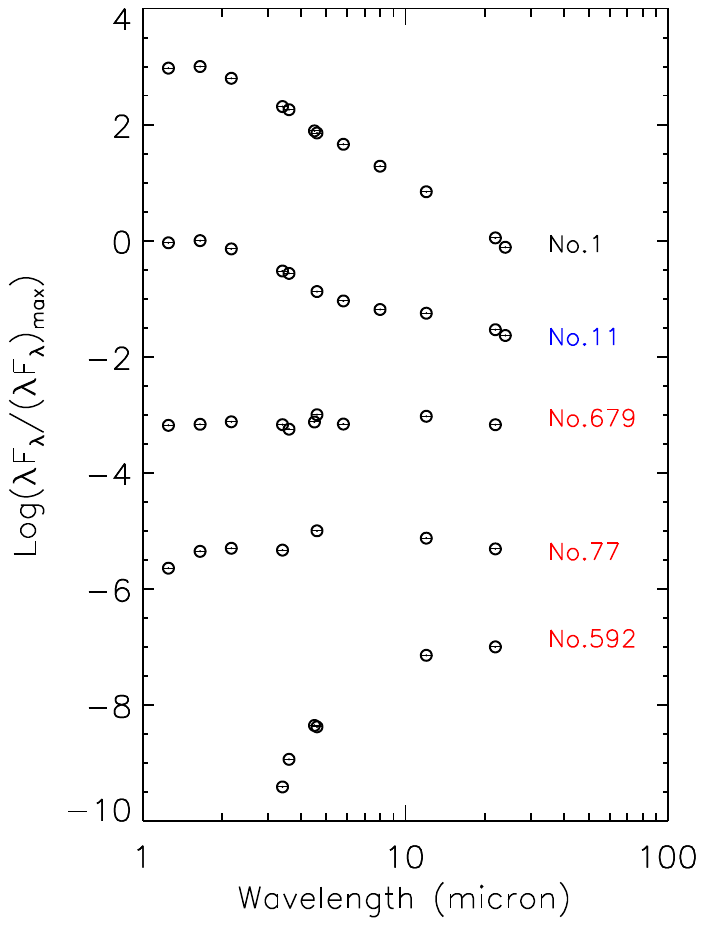}
\caption{Examples of SEDs with almost no IR excess (No.1), with IR excess (No.11), extreme IR excess (No.679, 77) and very extreme IR excess (No.592). SEDs have been normalized by their maximum fluxes and offset for display purposes. \label{sed_sample}}
\end{figure}

Besides the CMDs and CCDs, the spectral energy distribution (SED) could serve as another classification indicator. We note that since the intrinsic temperatures between RSGs and AGBs, even between them and the foreground dwarfs are not significantly different, plus the IR excess caused by the MLR in both RSGs and AGBs, we are only able to identify some extreme targets by using SED. Figure~\ref{sed_sample} shows several typical examples of SEDs with almost no IR excess (No.1), with IR excess (No.11), with extreme IR excess (No.679, 77) and with very extreme IR excess (No.592). Thus, Nos.77, 592, and 679 are considered as SED outliers and all of them have already been identified via the CMDs.

\subsection{Mid-infrared variability}

Except for the luminosity, color and SED, another way to separate RSGs from AGBs is by the variability. In the optical band, the typical amplitude of RSGs is about 0.5 to 1.0 magnitudes, which is much less than the AGBs \citep{Wood1983, Kiss2006, Groenewegen2009, Yang2011, Yang2012}. Thus, we expect the same in the MIR bands.

As mentioned before, since we used epoch-binned MIR time-series data from both ALLWISE and NEOWISE-R projects, there is a chance that the zero points of ALLWISE and NEOWISE-R may be different \footnote{http://wise2.ipac.caltech.edu/docs/release/neowise/expsup/\\sec2\_1c.html}. Therefore, the median absolute deviation (MAD) has been used to calculate the variability of each target, since it is mostly insensitive to outliers \citep{Rousseeuw1993, Sokolovsky2017}. MAD also can be converted to the standard deviation ($\sigma$), which is more intuitive, by using a constant scale factor of 1.4826 as $\sigma=MAD\times1.4826$ assuming a standard normal distribution. The median magnitude of the binned lightcurve has been adopted for each target as the average magnitude. Meanwhile, we also calculate the MAD within each epoch as an estimation of short-term (approximately five to ten days) variability.

\begin{figure*}
\center
\includegraphics[bb=55 400 555 695, scale=1.]{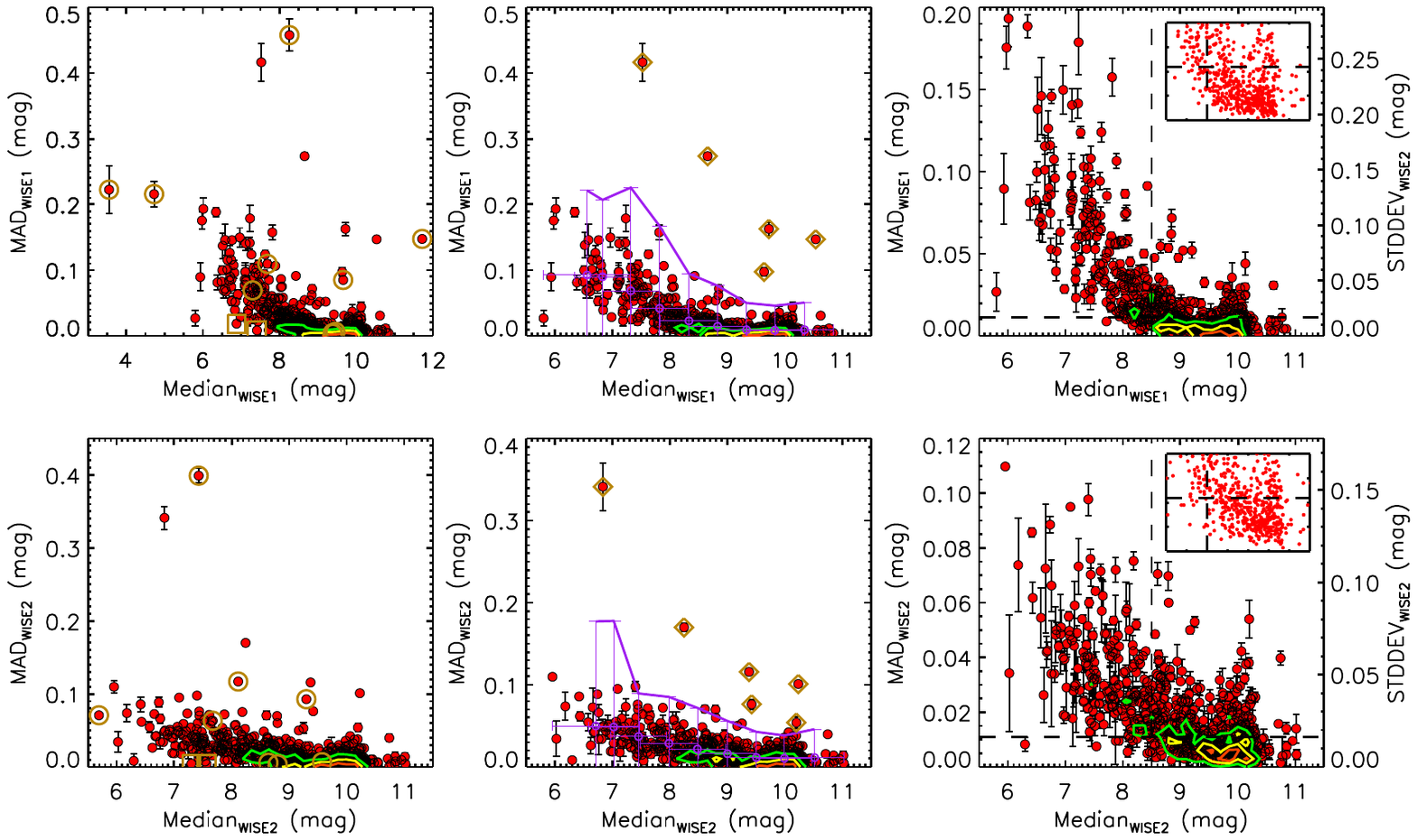}
\caption{WISE1/WISE2 median magnitude versus MAD diagrams. From left to right: initial sample with previous identified outliers; the same diagram with previous outliers removed and new outliers based on variability identified with dark yellow open diamonds; the final cleaned dataset with all outliers removed, where the insert panels show the range of $7.5<median_{WISE1/WISE2}<11.0~mag$ and $MAD_{WISE1/WISE2}<0.02~mag$. Error bars show the short-term (approximately five to ten days) variability within each epoch calculated by MAD (same below). See text for details. \label{mir_var}}
\end{figure*}

The left column of Figure~\ref{mir_var} shows the WISE1/WISE2 median magnitude versus MAD for the initial sample, with previously identified outliers indicated on the diagram. It can be seen that several outliers have already been identified by using CMDs and SEDs. The middle column of Figure~\ref{mir_var} shows the same diagrams with previous outliers removed. Still, there are few outliers deviating from the main population. Again, we use the method described in the CMD section to filter them out and mark them as dark yellow open diamonds. The variability outlier is defined when detected in both WISE1 and WISE2 bands. There are five of them, No.13, 221, 311, 346, 694. Finally, the right column of Figure~\ref{mir_var} shows the cleaned dataset of 744 targets in WISE1 and 746 targets in WISE2 bands. 

In total, we have 744 identified RSGs with both WISE1 and WISE2 time-series data which is $\sim96\%$ of the initial sample. We focus on this final sample in the following analysis. 

From the diagram, it can be seen that the MIR variability is increasing along with the increasing of brightness. The WISE2 band is more scattered than the WISE1 band which may be mainly due to the lower sensitivity of the longer wavelength. In total, $\sim$47\% of targets have $MAD_{WISE1}>0.011~mag$, which is at least three times larger than the robust sigma of photometric error ($\sim$0.0036 mag in both WISE1 and WISE2 band) and most likely indicates real variability considering the binned data we use (also see the insert panels of the right column of Figure~\ref{mir_var}), $\sim$31\% of targets have $median_{WISE1}<8.5~mag$, and $\sim$28\% of targets have both $MAD_{WISE1}>0.011~mag$ and $median_{WISE1}<8.5~mag$. Meanwhile, $\sim$57\% of targets have $MAD_{WISE2}>0.011~mag$, $\sim$27\% of targets have $median_{WISE2}<8.5~mag$, and $\sim$26\% of targets have both $MAD_{WISE2}>0.011~mag$ and $median_{WISE2}<8.5~mag$. About one quarter of the targets in the bright end (the brighter population) show evident variability compared to others (the fainter population). Here we notice that there may be a concern about the saturation problem, since the saturation limit for WISE1 band is $\sim8.0~mag$. However, since the WISE1 and WISE2 bands are showing the same trends while the saturation limit for WISE2 band is $\sim7.0~mag$, which indicates that the majority of targets are not saturated in WISE2, the trend must be real. Also, the typical fraction of saturated pixels in our magnitude range is $\lesssim10\%$ for $WISE1=7.0\sim8.0~mag$ and $\lesssim20\%$ for $W1=6.0\sim7.0~mag$. Since the WISE photometry is measured by using Point Spread Functions (PSFs), it gives a reasonable measurement when the majority of pixels are not saturated (the saturated pixels are extrapolated from the neighboring nonsaturated pixels). Even for the saturated pixels, in general, their values are likely to be underestimated but probably not overestimated in the photometry. Moreover, we have binned data within each epoch and used the MAD instead of standard deviation to measure the variability, which is more resistant to the outliers caused by saturation. Finally, we have also compared our results with the optical time-series data from ASAS-SN \citep{Shappee2014, Kochanek2017} and found that they are showing the similar tendencies (results will be presented in a future paper). Interestingly, we also notice that there is a peak ($MAD>0.011~mag$) around $median_{WISE1/WISE2}\sim10.0~mag$. Visual inspection indicates that most of the targets show long-term variation known as the long secondary period (LSP; \citealt{Olivier2003}). This may lead to a very interesting question about the origin of variability of RSGs (caused by pulsation, convection or even other reasons).

\begin{figure*}
\center
\includegraphics[bb=65 415 555 635, scale=0.8]{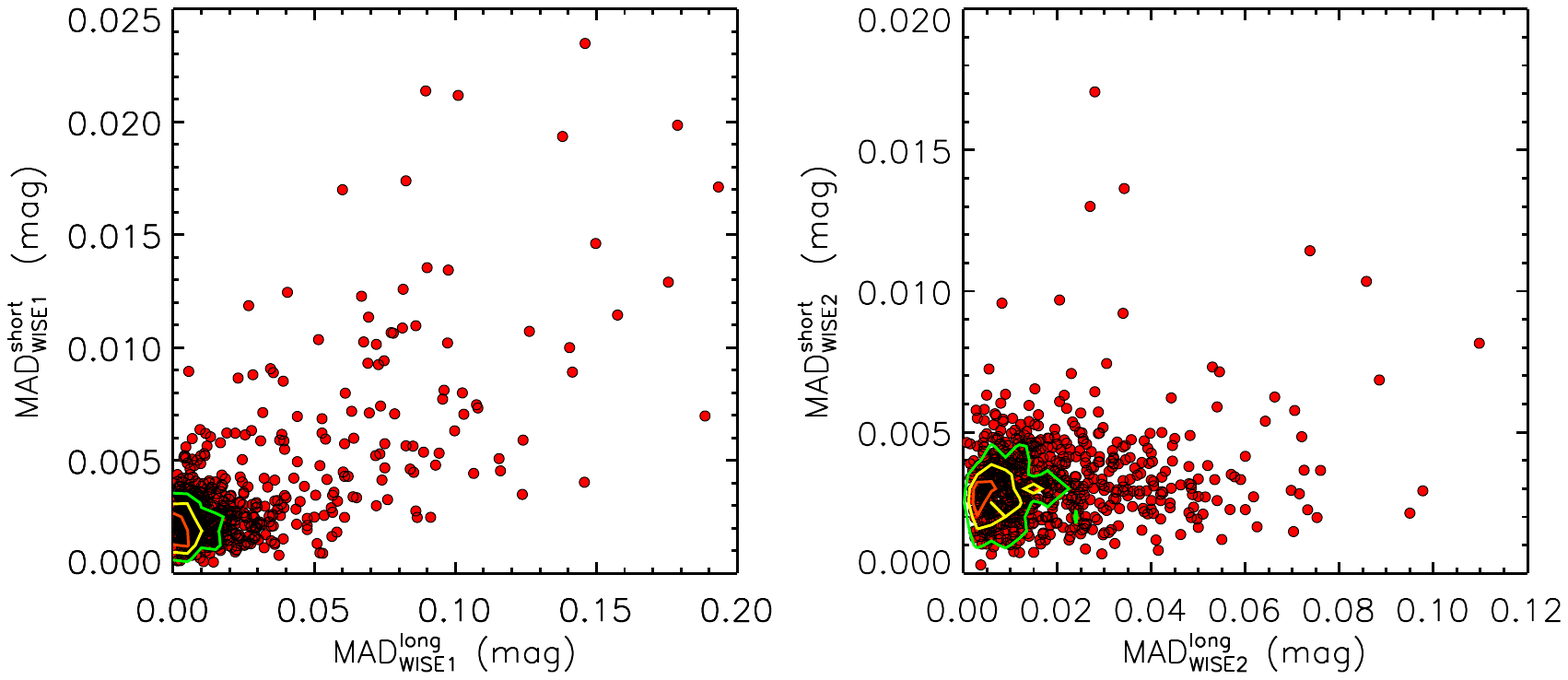}
\caption{Comparison of long- and short-term variabilities for the final sample. Both long- and short-term variabilities are increasing along with each other in WISE1 but not in WISE2 band, which may due to the smaller variability and lower sensitivity in the longer wavelength. \label{mir_var_ls}}
\end{figure*}

The long- and short-term variabilities also have been compared as shown in Figure~\ref{mir_var_ls}. There is a tendency in the WISE1 band that both long- and short-term variabilities are increasing along with each other. This may due to the short-term irregular variability appearing on the surface of the RSGs triggered by the pulsation or convection \citep{Stothers2010, Stothers2012}. This tendency is not observed in WISE2 band which may be due to the smaller variability and lower sensitivity in the longer wavelength.

\subsection{Outliers}

As we described above, CMD, SED, and variability are used to identify outliers in our sample. However, deviation from the main population may or may not indicate that those outliers are not RSGs. We have checked 13 outliers identified by CMD/SED (No.22, 77, 123, 204, 283, 291, 376, 394, 547, 592, 679, 723, and 773) and five outliers identified by variability (No.13, 221, 311, 346, and 694) through VizieR \citep{Ochsenbein2000}. Some peculiar outliers are discussed below.

\begin{itemize}[noitemsep,topsep=0pt,parsep=0pt,partopsep=0pt]

  \item No.22 is classified as an O-AGB by \citet{Kastner2008} based on the $JHK8$ color-based classification system, available literature, and additional crosschecks; a Mira-type O-AGB star based on the Optical Gravitational Lensing Experiment (OGLE)-III lightcurves by \citet{Soszyski2009} and \citet{He2016}, in which \citet{Soszyski2009} also indicate that a distinction between Miras and semiregular variables (SRVs) is not obvious for oxygen-rich stars; a far-infrared (FIR) object, whose SEDs rise from 8 to 24 $\mu$m~and typically are unresolved background galaxies, compact HII regions, planetary nebulae (PNe), or YSOs, based on the Spitzer-SAGE photometry by \citet{Boyer2011}; a Class I/II YSO candidate selected by support vector machine based on the 2MASS, WISE and Planck data by \citet{Marton2016}; a RSG based on the MIR spectrum from IRS by \citet{Jones2012, Jones2014, Jones2017}. 

  \item No.77 is a remarkable RSG, WOH G64, which has a thick circumstellar envelope, an unusual late spectral type, the largest radius ever known, a very low effective temperature, maser activity and nebular emission lines \citep{Westerlund1981, Elias1986, Wood1986, vanLoon2005, Ohnaka2008, Levesque2009}. Still, it is also classified as a Mira-type C-AGB star by \citet{Soszyski2009} and \citet{He2016}, a x-AGB based on the Spitzer-SAGE photometry by \citet{Srinivasan2009}, a Class I/II YSO candidate by \citet{Marton2016}.

  \item No.679 is classified as a ``blue'' star with $(B-V)_0<0.14$ and $M_{bol}<-7.0$ ($T_{eff}=4499$ and $M_{bol}=-7.84$) by \citet{Massey2002}; a luminous AGB based on the 2MASS color by \citet{Tsalmantza2006}; a RSG by \citet{Kastner2008}; a C-AGB by \citet{Srinivasan2009}; a C-AGB by \citet{Boyer2011}; a Class I/II YSO by \citet{Marton2016}; a RSG by \citet{Jones2012, Jones2014, Jones2017}.
  
  \item No.773 is the brightest target in our sample with $K_S\approx4.5~mag$ which is far brighter than the theoretical limit of 30$M_\sun$ ($K_S=5.77~mag$). The parallax of this target is $-0.80\pm0.62~mas$ with excess astrometric noise of $1.08~mas$ \citep{Gaia2016}, indicating a distance of $\sim1.25^{+4.305}_{-0.546}~kpc$. The proper motions of this target are $-7.241\pm1.471~mas/yr$ in right ascension and $16.810\pm1.712~mas/yr$ in declination. Considering also the large variation in the MIR bands (a full amplitude of $\sim1.0~mag$ in WISE1 band), the red color ($J-K_S\approx1.2$) and the similarity of MIR spectra between this target and O-AGBs, it is likely to be a foreground AGB star.

\end{itemize} 

\begin{figure*}
\center
\includegraphics[bb=65 425 530 635, scale=1.]{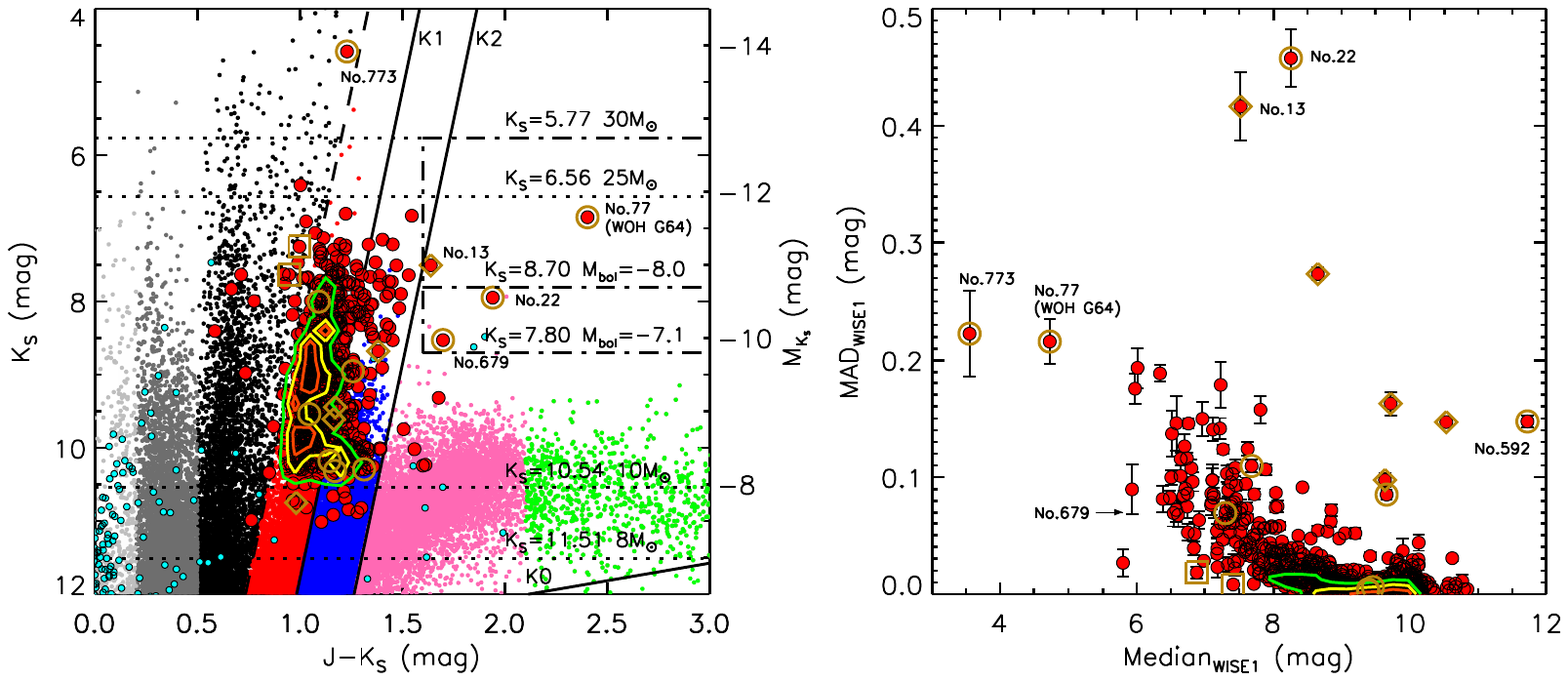}
\caption{Left: same as Figure~\ref{cmd_jhk} with peculiar outliers indicated on the diagram. Additional constrains of $J-K_S>1.6~mag$, $K_S<7.80~mag$ ($M_{bol}<-7.1~mag$), $K_S<8.70~mag$ ($M_{bol}<-8.0~mag$) and $K_S>5.77~mag$ ($<30M_\sun$) are shown by the dashed-dotted lines. Targets inside this region are likely to be either super-AGBs or RSGs. Right: same as Figure~\ref{mir_var} with peculiar outliers indicated. Considering luminosity, color and MIR variability from both panels, it appears that the No.13 and 22 might be super-AGBs while No.679 might be a dust-obscured RSG.
 \label{cmd_jhk_outlier}}
\end{figure*}

The left panel of Figure~\ref{cmd_jhk_outlier} shows the same diagram as Figure~\ref{cmd_jhk} with peculiar outliers indicated on the diagram. Additional constrains of $J-K_S>1.6~mag$, $K_S<7.80~mag$ ($M_{bol}<-7.1~mag$), $K_S<8.70~mag$ ($M_{bol}<-8.0~mag$) and $K_S>5.77~mag$ ($<30M_\sun$) are shown as the dashed-dotted lines. Targets inside this region are likely to be either super-AGBs or RSGs. The right panel of Figure~\ref{cmd_jhk_outlier} shows the same diagram as Figure~\ref{mir_var} with peculiar outliers indicated. Considering luminosity, color and MIR variability from both panels of Figure~\ref{cmd_jhk_outlier}, it seems like the No.13 and 22 may be super-AGBs while No.679 may be a dust-obscured RSG. 

\begin{figure*}
\center
\includegraphics[bb=65 430 560 675, scale=1.]{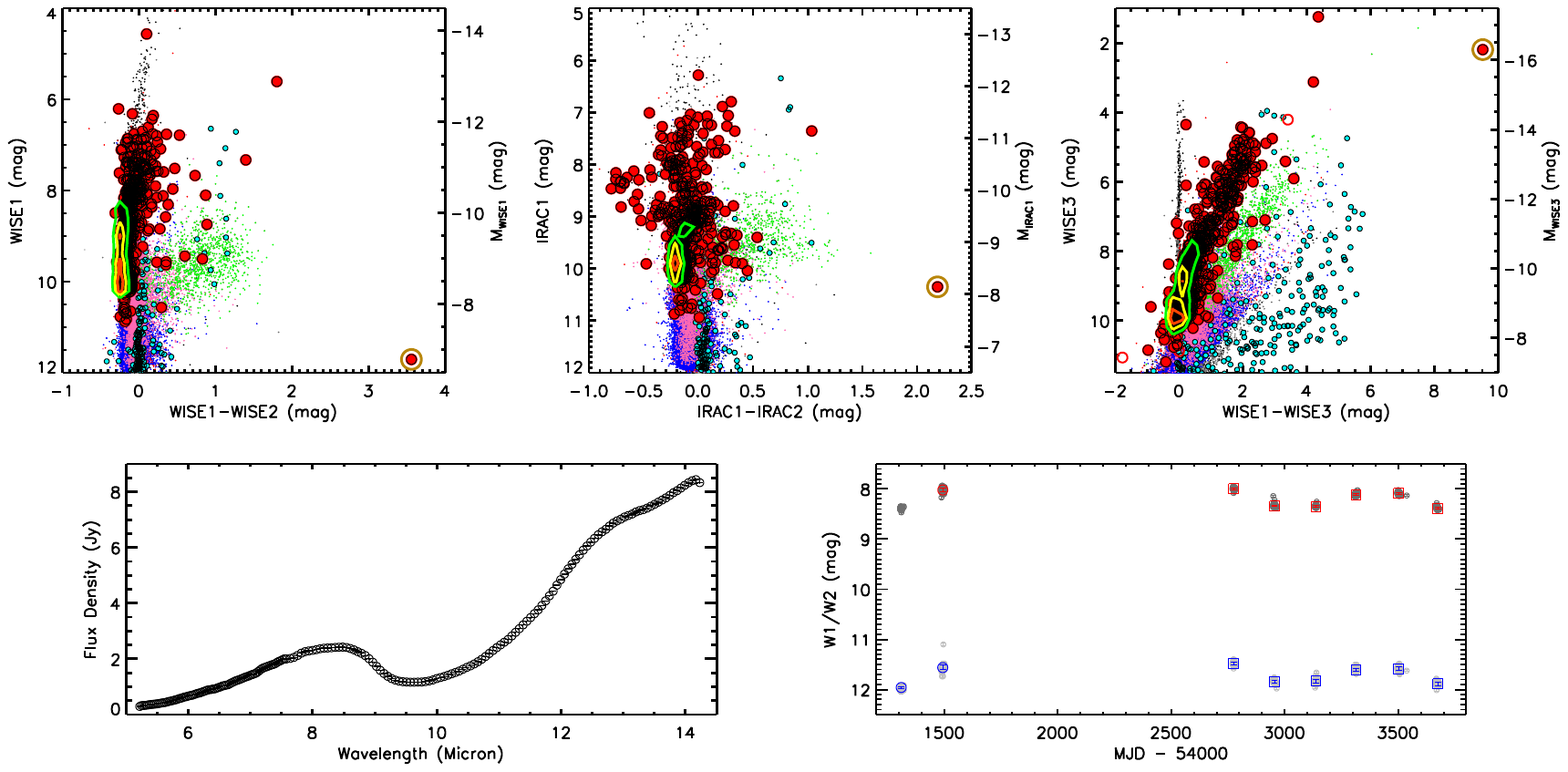}
\caption{Top panels: CMDs of the initial sample with the No.592 indicated. It is shown that the No.592 is the reddest target in our sample. An additional check without the $K_S$-band magnitude constrain ($K_S\leq12.5~mag$; Section 4.1) also shows that this target is the reddest object detected by WISE and Spitzer. Bottom left: Enhanced IRS spectrum of $\sim$5 to 14 $\mu$m. Bottom right: MIR lightcurves of WISE1 and WISE2 bands. See text for details. \label{n592}}
\end{figure*}

\begin{itemize}[noitemsep,topsep=0pt,parsep=0pt,partopsep=0pt]
  \item It seems to us that No.592 is the most interesting target in our sample. It is classified as a YSO based on the IR excess from Spitzer data by \citet{Whitney2008} and  \citet{Gruendl2009}; a x-AGB by \citet{Srinivasan2009} and \citet{Vijh2009}; a RSG by \citet{Jones2012, Jones2014, Jones2017}. The nonRSG classifications are mainly based on the extreme MIR color (upper panels of Figure~\ref{n592}), since it is not visible in the optical and NIR bands. An additional check without the $K_S$-band magnitude constrain ($K_S\leq12.5~mag$; Section 4.1) also shows that this target is the reddest object detected by WISE and one of the reddest objects detected by Spitzer. However, if the MIR spectral classification is reliable (no redshift feature, no featureless continuum with PAHs or atomic emission lines, solely molecular features, no ice absorption at 15 $\mu$m, no C-rich features of 5.0, 7.5, and 13.7 $\mu$m~absorption, $M_{bol}<-7.1~mag$ and SED peak $\sim$1 $\mu$m; \citealt{Jones2017}; see the enhanced IRS spectrum of $\sim$5 to 14 $\mu$m~at the bottom left panel of Figure~\ref{n592}), then probably we are witnessing a RSG right before the explosion, since the MLR of RSGs just before the explosions could be as high as $\sim10^{-4}$ to $10^{-2}~M_\sun~yr^{-1}$ \citep{Moriya2011, Moriya2015}, which may produce a very thick dust envelope and highly obscure RSGs in the optical and NIR bands. It is also possible that some of the YSO and x-AGB candidates are pre-explosion RSGs which may be difficult to be confirmed by the optical or even NIR spectroscopy due to the heavily-obscured environment. This scenario may be also related to the intermediate luminosity optical transients (ILOTs; \citealt{Prieto2008, Bond2009, Berger2009}) or the Object X in M33 \citep{Khan2011, Mikolajewska2015}. Other options for this target would be a x-AGB in the very late evolutionary stage \citep{Boyer2015} based on the reddest color in the LMC and the semiregular lightcurve. In any case, this target is quite interesting and needs to be followed up with deep optical-to-NIR spectroscopic observations.
\end{itemize}

Based on the above analysis, we emphasize that there is an uncertainty for the classification of RSGs based on the brightness and color in optical/NIR bands, since some of them may be highly reddened or obscured by the surrounding dust envelope and not even seen in the NIR, especially for the RSGs in the very late evolutionary stage. Also, due to different spectral features captured by different filters in the IR bands, some RSGs may show different characteristics than the others. High resolution or deep spectra, longer wavelength photometry and additional variability information may help to confirm those targets.

Our sample of 744 verified RSGs is almost four times larger than the sample of Paper \uppercase\expandafter{\romannumeral1}. Still, this only represents a small percentage (presumably 10$\sim$15\%) of RSGs in the LMC predicted by the theoretical cuts (see Figure~\ref{cmd_jhk}) and further large-scale investigation is strongly needed to fully understand the nature of RSGs.

\section{Discussion}

\subsection{Variability, luminosity, and mass loss rate}

\begin{figure*}
\center
\includegraphics[bb=55 410 558 665]{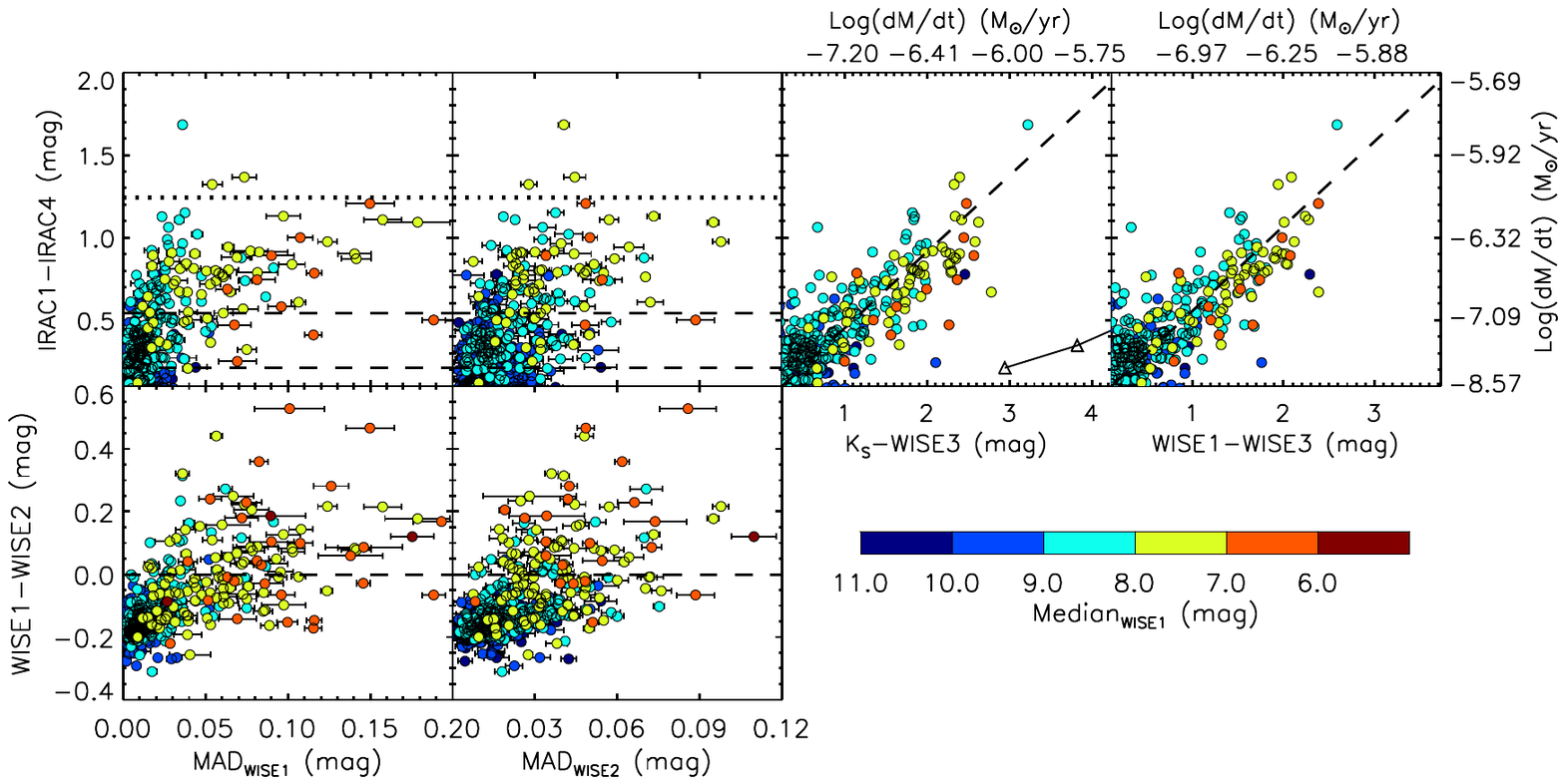}
\caption{Variability versus infrared color diagrams. Upper left: $MAD_{{WISE1}/{WISE2}}$ versus $IRAC1-IRAC4$. The MLR is calculated by using the formula from \citet{Groenewegen2018}, which is increasing along with the variability and probably approaching an upper limit of $\sim-6.1~M_\sun/yr^{-1}$ as shown by the dotted lines. The limits of $MLR=-8.0~M_\sun/yr^{-1}$ and $-7.0~M_\sun/yr^{-1}$ are also shown by the dashed lines. Upper right: a linear relation between $IRAC1-IRAC4$ and $K_S-WISE3$/$WISE1-WISE3$ which can be used to derive MLR in $K_S-WISE3$ and $WISE1-WISE3$ colors with slightly larger error. For comparison, the MLR calculated by using $K-[12]$ color based on the MIR data from IRAS \citep{Josselin2000} has been added in the $K_S-WISE3$/$IRAC1-IRAC4$ diagram as shown by the open triangles ($MLR_{K-[12]}=-8.0,~-7.5~M_\sun/yr^{-1}$). It can be seen that the MLR calculated by using $K-[12]$ is largely underestimated compared with the one calculated by using $IRAC1-IRAC4$. Bottom: Previous known ``blue'' RSGs ($WISE1-WISE2$/$IRAC1-IRAC2<0$) in $WISE1-WISE2$/$IRAC1-IRAC2$ are turning redder ($WISE1-WISE2$/$IRAC1-IRAC2>0$) along with the increasing of variability. \label{mir_color_var}}
\end{figure*}

An important issue about RSGs is the MLR. For the nearby Universe, the main contribution of the dust comes from low-mass AGBs and supernovae (SNe). But RSGs may be a major contributor at high redshift along with the Type II SNe, due to the absence of old low-mass AGB stars. In a recent study, \citet{Groenewegen2018} investigate MLR and luminosity in a large sample of evolved stars in several Local Group galaxies with a variety of metallicities and star-formation histories based on the MIR spectra from IRS and the complementary optical and infrared photometry. Their results show a correlation between MLR and $IRAC1-IRAC4$ color for M-type stars as
\begin{equation}
log\dot{M}=-4.078-\frac{2.488}{(IRAC1-IRAC4)+0.545}.
\end{equation}
This relation is only valid when $IRAC1-IRAC4\geq0.1~mag$ due to the quick drop of MLR in the bluest color. Since we find a positive correlation between luminosity and variability (Figure~\ref{mir_var}), and there is a positive correlation between MLR and luminosity \citep{Verhoelst2009, Bonanos2010, Mauron2011}, it would be natural to expect that the variability is also related to the MLR, e.g., targets with larger variability also show larger MLR. The two panels in the upper left of Figure~\ref{mir_color_var} show the $MAD_{{WISE1}/{WISE2}}$ versus $IRAC1-IRAC4$. The MLR is calculated by using the above formula. As expected, the MLR is increasing along with the variability and probably approaching an upper limit of $\sim-6.1~M_\sun/yr^{-1}$ as shown by the dotted lines. The limits of $MLR=-8.0~M_\sun/yr^{-1}$ and $-7.0~M_\sun/yr^{-1}$ are also shown by the dashed lines. This trend is more obvious in WISE1 than in WISE2 band which may be due to the lower sensitivity and smaller variability in WISE2 band. In total, there are 438 valid targets ($IRAC1-IRAC4\geq0.1$) in both IRAC1 and IRAC4 bands, of which $\sim0.7\%$ (three targets) of them have $MLR>-6.1~M_\sun/yr^{-1}$, $\sim23\%$ (100 targets) of them have $-7.0~M_\sun/yr^{-1}<MLR\leq-6.1~M_\sun/yr^{-1}$, $\sim41\%$ (178 targets) of them have $-8.0~M_\sun/yr^{-1}<MLR\leq-7.0~M_\sun/yr^{-1}$. 

Meanwhile, since there is a linear relation between $IRAC1-IRAC4$ and $K_S-WISE3$/$WISE1-WISE3$ as shown by the two panels in the upper right of Figure~\ref{mir_color_var}, the MLR also can be derived by using those two colors with slightly larger error but benefiting from the full sky coverage of 2MASS and WISE as 
\begin{equation}
log\dot{M}=-4.078-\frac{2.488}{0.463\times(K_S-WISE3)+0.534},
\end{equation}
\begin{equation}
log\dot{M}=-4.078-\frac{2.488}{0.513\times(WISE1-WISE3)+0.587},
\end{equation}
which are valid for $K_S-WISE3\geq0.24~mag$ and $WISE1-WISE3\geq0.11~mag$, respectively. As stated by \citet{Groenewegen2018}, the estimated MLR can be different from previous studies due to differences in adopted optical constants and details in the radiative transfer modeling. For comparison, the MLR calculated by using $K-[12]$ color based on the MIR data from IRAS \citep{Josselin2000} has been added in the $K_S-WISE3$/$IRAC1-IRAC4$ diagram as shown by the open triangles ($MLR_{K-[12]}=-8.0,~-7.5~M_\sun/yr^{-1}$), when taking into account of the difference between K and $K_S$ ($K=K_S+0.044$; \citealt{Carpenter2001}) and the difference between IRAS [12]-band and WISE3 ($[12]=WISE3-0.435$; \citealt{Dorda2016}). It can be seen that the MLR calculated by using $K-[12]$ is largely underestimated compared with the one calculated by using $IRAC1-IRAC4$. 

Moreover, RSGs are known to be ``blue'' in $WISE1-WISE2$/$IRAC1-IRAC2$ color, which is mainly caused by the CO absorption. However, it also appears that RSGs are turning into redder color along with the increasing of variability as shown in the bottom panels of Figure~\ref{mir_color_var}.

\begin{figure*}
\center
\includegraphics[bb=70 360 540 720, scale=1.]{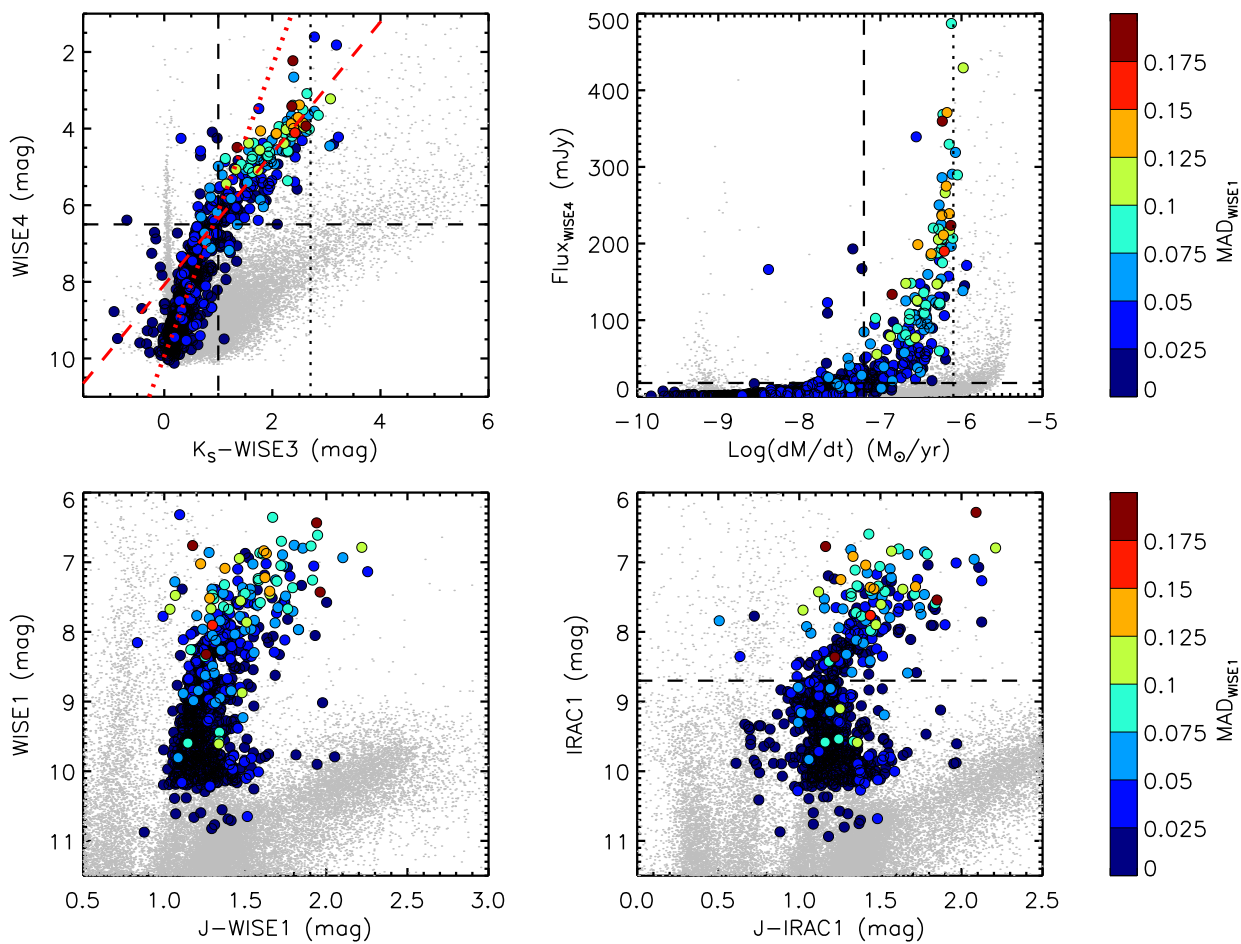}
\caption{CMDs with color coded MIR variability. Upper left panel: WISE4 magnitude versus $K_S-WISE3$ color. The MIR variability is evident ($median_{MAD_{WISE1}}\approx0.048~mag$) for targets with $K_S-WISE3\geq1.0~mag$ and $WISE4\leq6.5~mag$, while targets with $K_S-WISE3\leq1.0~mag$ and $WISE4\geq6.5~mag$ exhibit much smaller MIR variability ($median_{MAD_{WISE1}}\approx0.007~mag$). Upper right panel: WISE4 flux versus MLR. The WISE4 (22 $\mu$m) flux grows exponentially along with the increasing of the volume or brightness of the star. Since the variability and luminosity are degenerate, both of them may have important contribution to the MLR. Lower panels: WISE1/$J-WISE1$ and IRAC1/$J-IRAC1$ diagrams. A turning point around $IRAC1\approx8.7~mag$ is shown by the dashed line in IRAC1/$J-IRAC1$ diagram separates well two populations (brighter and fainter) of RSGs with different MIR variability. \label{mir_cmd_var}}
\end{figure*}

For further information on the relation between variability, luminosity and MLR, the combination of variability and CMDs has been investigated. Due to the strict cuts on SEIP source list, there are much fewer detection in MIPS24 than the WISE4 bands (554 versus 742 targets). Thus, we show the WISE4 magnitude and flux versus $K_S-WISE3$/MLR along with the color coded MIR variability in the upper panels of Figure~\ref{mir_cmd_var}. It can be seen that there is a relatively tight correlation between the MIR variability, MLR (in terms of $K_S-WISE3$ color) and the warm dust or continuum (in terms of WISE4 magnitude or flux), where the MIR variability is evident ($median_{MAD_{WISE1}}\approx0.048~mag$) for targets with $K_S-WISE3\geq1.0~mag$ and $WISE4\leq6.5~mag$, while targets with $K_S-WISE3\leq1.0~mag$ and $WISE4\geq6.5~mag$ are showing much smaller MIR variability ($median_{MAD_{WISE1}}\approx0.007~mag$). Two robust linear functions are fitted for targets with $K_S-WISE3\geq1.0~mag$ and $WISE4\leq6.5~mag$ (red dashed line), and targets with $K_S-W3\leq1.0~mag$ and $WISE4\geq6.5~mag$ (red dotted line) as,
\begin{equation}
WISE4=-1.72\times(K_S-WISE3)+8.08,
\end{equation}
\begin{equation}
WISE4=-3.79\times(K_S-WISE3)+9.95.
\end{equation}
Even if there is a upper limit of $-6.1~M_\sun/yr^{-1}$ for the MLR per unit volume as shown by the dotted line, the WISE4 (22 $\mu$m) flux still grows exponentially along with the increasing of the volume or brightness of the star. However, since the variability also increases with the luminosity, they are degenerate and both of them may have important contributions to the MLR. 

Meanwhile, the two populations (brighter and fainter) of RSGs with different MIR variabilities are also distinguished in the $WISE1/J-WISE1$ and $IRAC1/J-IRAC1$ diagrams with the larger variability population being redder and brighter as shown in the lower panels of Figure~\ref{mir_cmd_var}. Especially, the difference is more obvious in the IRAC1/J-IRAC1 diagram in which a turning point appears around $IRAC1\approx8.7~mag$ as shown by the dashed line compared with the $WISE1/J-WISE1$ diagram. This may be due to different sensitivities, effective wavelengths, and bandwidths between the IRAC1 and WISE1 bands which capture different spectral features. 

The combination of Figure~\ref{mir_cmd_var} and the diagrams from Appendix clearly shows a significant difference between the brighter and fainter populations of RSGs in the relatively longer wavelengths with the brighter one showing much larger infrared excess, which is likely result from the excess circumstellar dust caused by larger MLR of the brighter population. In particular, this difference is not seen before, since the majority of previous studies only focus on the brighter population, while the fainter population has not been revealed until recent spectroscopic or photometric surveys. However, up till now, it is not clear which mechanism is mainly responsible for the infrared excess of RSGs, for example, episodic mass loss, dust-driven stellar winds, pulsation, convection, luminosity, metallicity, binarity, rotation, dissipation of Alfv\`en waves, or the combination of several factors \citep{Macgregor1992, Harper2001, Levesque2007, Massey2007, Bonanos2010, Yoon2010, Beasor2016}. A relatively complete sample of RSGs including both bright and faint end of magnitude at different metallicities would be more helpful to understand the evolution and mass loss of RSGs, and the important role they play in the massive star evolution.

\subsection{Variability, extinction and evolutionary model}

\begin{figure*}
\center
\includegraphics[bb=150 380 450 710, scale=1.]{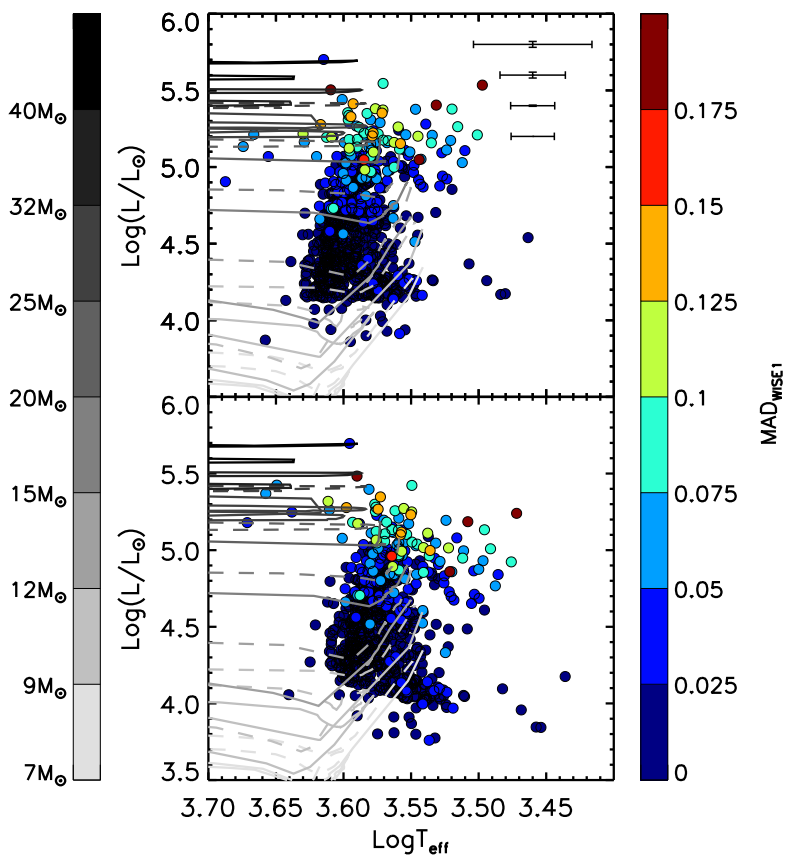}
\caption{Hertzsprung-Russell diagram overlapped with Geneva stellar evolutionary tracks at solar metallicity of $Z=0.014$. The nonrotation (solid lines) and rotation ($V/V_C=0.40$; dashed lines) models of 7 to $40~M_\sun$ are color coded. Top: luminosity is converted by using a constant bolometric correction of $BC_{K_S}=2.69$, while $T_{eff}$ is derived following the method used in \citet{Neugent2012}. Most of the outliers can be well explained by the combination of variability and reddening as shown by four error bars. Bottom: luminosity and $T_{eff}$ are converted by using the exact same method used in \citet{Neugent2012}. See text for details. \label{geneva}}
\end{figure*}

Finally, we also compared our result with the Geneva stellar evolutionary tracks at solar metallicity of $Z=0.014$ as shown in Figure~\ref{geneva}. The nonrotation (solid lines) and rotation ($V/V_C=0.40$; dashed lines) models of 7 to $40~M_\sun$ are color coded in the diagram \citep{Ekstrom2012}. Two different sets of parameters were used to convert the $K_S$-band magnitude and $J-K_S$ color to the luminosity and effective temperature. For the upper panel of Figure~\ref{geneva}, the luminosity is converted from $K_S$-band magnitude by using a constant bolometric correction of $BC_{K_S}=2.69$ \citep{Davies2013}, since the BC of RSGs is independent of IR color and the uncertainty of $BC_{K_S}$ is about $15\%\sim25\%$ \citep{Buchanan2006, Davies2013}. The $T_{eff}$ is derived following the method used in \citet{Neugent2012}. Here we adopted a larger color excess for de-reddening with $A_{K_S}=0.1~mag$, $A_J/A_{K_S}=2.72$ and $E(J-K_S)=0.172$ which is equivalent to $A_V\approx1.0~mag$ \citep{Gao2013, Xue2016}, than the one used in \citet{Neugent2012} ($E(J-K_S)=0.07$). This is mainly due to two reasons. First, it can be seen from Figure~\ref{spatial} that most of our targets are close to the star formation region where higher reddening is expected even in the NIR bands. The values we used here may still underestimate the actual extinction and reddening. Second, the reddening used in \citet{Neugent2012} ($E(J-K_S)=0.07$) is appropriate for the early-type stars but probably underestimated for RSGs \citep{Massey2005, Levesque2005, Levesque2006}. For comparison, the lower panel of Figure~\ref{geneva} shows the conversion by using the exact same method used in \citet{Neugent2012} with the smaller reddening. Meanwhile, the BC proposed by \citet{Bessell1984} also has been investigated, which gives a very similar result as \citet{Neugent2012}. It can be seen that the majority of targets in the upper panel are following the tracks now by using a larger extinction, compared to the lower panel. Most of the outliers in the upper panel also can be explained by the combination of variability and reddening as shown by four error bars. The main reason is that the J and $K_S$ magnitudes from 2MASS are single-epoch ``snapshots'' rather than the average magnitudes, in which $J-K_S$ may underestimate or overestimate the colors of RSGs, especially for the targets with larger variability. This also can be partially confirmed by few targets found in the 2MASS 6X Point Source Working Database \citep{Cutri2012} with multiple observations in which $J-K_S$ varies more than 0.1 mag. Here we assume that the same level of variability is adopted for $J-K_S$ as well as WISE1 band, which is a good approximation but may be still not enough. The first error bar indicates a large variability ($mean_{MAD_{J-K_S}}\approx0.05~mag$, equal to the mean $MAD_{WISE1}$ with $median_{WISE1}<8.5~mag$) and a large extinction ($E(J-K_S)=0.172$). The second one indicates a large variability and a small extinction ($E(J-K_S)=0.07$). The third one indicates a small variability ($mean_{MAD_{J-K_S}}\approx0.01~mag$, equal to the mean $MAD_{WISE1}$ with $median_{WISE1}\geq8.5~mag$) and a small extinction. The last one indicates a small extinction without variability. As the error bars show, the outliers in the bright end can be well explained by the combination of variability and reddening, while few outliers in the faint end may still need to be investigated.

\section{Summary}

We present a comprehensive investigation of IR properties and MIR variability of the largest sample of RSGs in the LMC up to now. The initial sample of 773 RSGs candidates are compiled from the literature. The co-added infrared data are collected from the SEIP source list, which contains 12 NIR to MIR bands data from 2MASS, Spitzer and WISE. The $\sim$6.6-year  MIR time-series data are collected from both ALLWISE and NEOWISE-R projects by applying several cut-offs and binned according to the observational epochs.

The initial sample is verified by using CMDs which exclude the most frequent outliers (No.22, 77, 204, 283, 291, 376, 394, 592, 679, 723, 773) and foreground targets (No.123, 547), SEDs which identify the extreme targets (No.77, 592, 679), and MIR variabilities which are assumed to separate the RSGs from AGBs (No.13, 221, 311, 346, 694). Based on the CCDs, it indicates that about 15\% of valid targets in the $IRAC1-IRAC2$/$IRAC2-IRAC3$ diagram may show PAH emission feature. We also show that arbitrary dereddening Q parameters related to the IRAC4, S9W, WISE3, WISE4 and MIPS24 bands could be constructed based on a precise measurement of the MIR interstellar extinction law. 

Several peculiar outliers in our sample have been discussed. By considering luminosity, color and MIR variability, it seems like the No.13 and 22 may be super-AGBs, while No.679 may be a dust-obscured RSG. Particularly, No.592 may be a RSG right before the explosion or a x-AGB star in the very late evolutionary stage based on the MIR spectrum and photometry. 

We emphasize that there is an uncertainty in the classification of RSGs based on the brightness and color in optical or NIR bands, since some of them may be highly reddened or obscured by the surrounding dust envelope and not even seen in the NIR, especially for RSGs in the very late evolutionary stage. Also, due to different spectral features captured by different filters in the IR bands, some RSGs may show different characteristics than the others. High resolution or deep spectra, longer wavelength photometry and additional variability information may help to confirm those targets.

After removing outliers, there are 744 identified RSGs having both the WISE1- and WISE2-band time-series data. The results show that the MIR variability is increasing along with the increasing brightness. In total, $\sim$47\% of targets have $MAD_{WISE1}>0.011~mag$, $\sim$31\% of targets have $median_{WISE1}<8.5~mag$, and $\sim$28\% of targets have both $MAD_{WISE1}>0.011~mag$ and $median_{WISE1}<8.5~mag$. Meanwhile, there are $\sim$57\% of targets have $MAD_{WISE2}>0.011~mag$, $\sim$27\% of targets have $median_{WISE2}<8.5~mag$, and $\sim$26\% of targets have both $MAD_{WISE2}>0.011~mag$ and $median_{WISE2}<8.5~mag$. About one quarter of targets in the bright end show evident variability compared to others. 

We also show that there is a relatively tight correlation between the MIR variability, MLR which is calculated by using the formula from \citet{Groenewegen2018} and converted to $K_S-WISE3$ color, and the warm dust or continuum (in terms of WISE4 magnitude or flux), where the MIR variability is evident ($median_{MAD_{WISE1}}\approx0.048~mag$) for targets with $K_S-WISE3>1.0~mag$ and $WISE4<6.5~mag$, while the rest of the targets show much smaller MIR variability. The MIR variability is also correlated with the MLR for which targets with larger variability also showing larger MLR with an approximate upper limit of $-6.1~M_\sun/yr^{-1}$. The variability and luminosity may be both important for the MLR since the WISE4-band flux is increasing along with the degeneracy of luminosity and variability. 

The combination of luminosity, color, variability and MLR shows a significant difference between two populations of RSGs with the brighter population showing much larger infrared excess (MLR) and variability compared to the fainter one. However, the main reason for this difference is still unclear since several factors are degenerated. Thus, a relatively complete sample of RSGs including both bright and faint end of magnitude at different metallicities would be more helpful to understand the evolution and mass loss of RSGs, and the important role they play in the massive star evolution.

The sample of identified RSGs has been compared with the Geneva evolutionary models. It is shown that the discrepancy between observation and evolutionary models can be mitigated by considering both variability and extinction.

\section{Acknowledgments}

The authors would like to thank the anonymous referee for many constructive comments and suggestions. This study is supported by the Hubble Catalog of Variables project, which is funded by the European Space Agency (ESA) under contract No.4000112940. This study is also partly supported the National Natural Science Foundation of China (Grant No.U1631104).

This publication makes use of data products from the Two Micron All Sky Survey, which is a joint project of the University of Massachusetts and the Infrared Processing and Analysis Center/California Institute of Technology, funded by the National Aeronautics and Space Administration and the National Science Foundation.
This work is based in part on observations made with the Spitzer Space Telescope, which is operated by the Jet Propulsion Laboratory, California Institute of Technology under a contract with NASA.

This publication makes use of data products from the Wide-field Infrared Survey Explorer, which is a joint project of the University of California, Los Angeles, and the Jet Propulsion Laboratory/California Institute of Technology. It is funded by the National Aeronautics and Space Administration.

This publication makes use of data products from the Near-Earth Object Wide-field Infrared Survey Explorer (NEOWISE), which is a project of the Jet Propulsion Laboratory/California Institute of Technology. NEOWISE is funded by the National Aeronautics and Space Administration.

This research has made use of the VizieR catalog access tool, CDS, Strasbourg, France.
This research has made use of the Tool for OPerations on Catalogues And Tables (TOPCAT; \citealt{Taylor2005}).

\begin{table*}
\caption{Initial sample of 773 RSG candidates} 
\label{isample}
\centering
\begin{tabular}{cccccccccc}
\toprule\toprule
ID & R.A.(J2000) & Decl.(J2000) & 2MASS\_J & 2MASS\_H & ... & $MAD^{short}_{WISE2}$ & $AMP_{WISE2}$ & Comment\tablefootmark{a} & Reference\tablefootmark{b}\\
 & (deg) & (deg) & (mag) & (mag) & ... & (mag) & (mag) &  &  \\
\midrule 
 1 &  70.92875  &   -67.782083 &   10.895 &  10.062 &  ...  &  0.0022  &  0.045  &  & N12 \\
 2 &  71.47275  &   -67.552833 &   10.948 &  10.097 &  ...  &  0.0024  &  0.025  &  & N12 \\
 3 &  71.6685   &   -67.186056 &   10.741 &   9.928 &  ...  &  0.0014  &  0.036  &  & N12 \\
 4 &  71.8277   &   -69.7057   &   10.114 &   9.292 &  ...  &  0.0037  &  0.062  &  & J17 \\
 5 &  72.00029  &   -68.819778 &   11.004 &  10.133 &  ...  &  0.0031  &  0.024  &  & N12 \\
...  &   ...   &  ...  &  ...  &  ...    &  ...    &   ...  &  ...     &   ...    &  ...\\
\midrule 
\end{tabular}
\tablefoot{
This table is available in its entirety in CDS. A portion is shown here for guidance regarding its form and content.\\
\tablefoottext{a}{C: CMD outlier, F: Foreground outlier, S: SED outlier, V: Variability outlier.}
\tablefoottext{b}{M03: \citet{Massey2003}, Y11: \citet{Yang2011}, N12: \citet{Neugent2012}, G15: \citet{Gonzalez2015}, J17: \citet{Jones2017}.}
}
\end{table*}

\begin{table*}
\caption{General information of co-added infrared data for the initial sample} 
\label{tbl1}
\begin{tabular}{cccccc}
\toprule\toprule
Band & Wavelength & Detected targets & Valid targets & $N_{Valid}/N_{Detected}$ & $N_{Valid}/N_{Total}$ \\
         & ($\mu$m) & & ($S/N\geq2$) & & \\
\midrule 
2MASS\_J & 1.25 & 772 & 772 & 100\% & 99.9\% \\
2MASS\_H & 1.65 & 772 & 772 & 100\% & 99.9\% \\
2MASS\_$K_S$ & 2.17 & 772 & 772 & 100\% & 99.9\% \\
IRAC1 & 3.6 & 692 & 692 & 100\% & 89.5\% \\
IRAC2 & 4.5 & 689 & 689 & 100\% & 89.1\% \\
IRAC3 & 5.8 & 706 & 706 & 100\% & 91.3\% \\
IRAC4 & 8.0 & 697 & 697 & 100\% & 90.2\% \\
MIPS24 & 24 & 570 & 570 & 100\% & 73.7\% \\
WISE1 & 3.4 & 767 & 767 & 100\% & 99.2\% \\
WISE2 & 4.6 & 767 & 767 & 100\% & 99.2\% \\
WISE3 & 12 & 767 & 761 & 99.2\% & 98.4\% \\
WISE4 & 22 & 766 & 623 & 81.3\% & 80.6\% \\
\midrule 
\end{tabular}
\end{table*}

\begin{table*}
\caption{Observation epochs of ALLWISE and NEOWISE-R} 
\label{tbl2}
\begin{tabular}{cc}
\toprule\toprule
Beginning & Ending \\
(MJD-54000) & (MJD-54000) \\
\midrule
ALLWISE & \\
\midrule
1200 & 1400 \\
1400 & 1600 \\
\midrule
NEOWISE-R & \\
\midrule
2600 & 2700 \\
2700 & 2885 \\
2885 & 3065 \\
3065 & 3245 \\
3245 & 3430 \\
3430 & 3610 \\
3610 &  \\
\midrule
\end{tabular}
\end{table*}

\begin{table*}
\caption{Outliers identified in $K_S$-based CMDs} 
\label{ks_outlier}
\begin{tabular}{cc}
\toprule\toprule
CMDs & IDs \\
\midrule
$K_S/K_S-WISE1$ & 22,44,77,181,204,291,311,376,400,445,474,679,723\\
$K_S/K_S-IRAC1$ & 69,89,204,291,376,394,473,543,549,679,723\\
$K_S/K_S-IRAC2$ & 13,204,291,549,566,679\\
$K_S/K_S-WISE2$ & 77,204,291,376,445,549,565,679,723\\
$K_S/K_S-IRAC3$ & 204,291,376,679,723\\
$K_S/K_S-IRAC4$ & 291,376,400,445,565,723\\
$K_S/K_S-WISE3$ & 77,291,376,445,649,672,679,723\\
$K_S/K_S-WISE4$ & 50,77,217,303,376,547,649,678,679,694,763\\
$K_S/K_S-MIPS24$ & 123,204,215,225,291,376,547,723\\
\midrule
\end{tabular}
\end{table*}

\begin{table*}
\caption{Outliers identified in WISE1-based CMDs} 
\label{w1_outlier}
\begin{tabular}{cc}
\toprule\toprule
CMDs & IDs \\
\midrule
$WISE1/J-WISE1$ & 77,181,204,217,283,291,376,474,555,644,649,679\\
$WISE1/H-WISE1$ & 22,77,181,204,217,283,291,376,400,412,439,474,644,649,679,723\\
$WISE1/K_S-WISE1$ & 22,44,77,181,204,217,283,291,311,376,412,439,474,649,656,679,723\\
$WISE1/WISE1-IRAC1$ & 22,62,69,181,283,394,493\\
$WISE1/WISE1-IRAC2$ & 13,69,291,311,346,566,679,694\\
$WISE1/WISE1-WISE2$ & 13,22,77,204,221,283,291,311,320,346,376,679,694,723\\
$WISE1/WISE1-IRAC3$ & 13,22,204,291,311,346,376,474,679,723\\
$WISE1/WISE1-IRAC4$ & 291,311,346,376,636,723\\
$WISE1/WISE1-WISE3$ & 22,77,204,217,225,291,325,376,606,636,649,672,679,694\\
$WISE1/WISE1-WISE4$ & 22,50,77,123,217,303,306,376,547,636,649,658,678,679,694,763\\
$WISE1/WISE1-MIPS24$ & 22,123,204,215,225,291,376,547\\
\midrule
\end{tabular}
\end{table*}

\begin{table*}
\caption{Outliers identified in IRAC1-based CMDs} 
\label{i1_outlier}
\begin{tabular}{cc}
\toprule\toprule
CMDs & IDs \\
\midrule
$IRAC1/J-IRAC1$ & 62,204,217,291,376,394,555,644,649,679\\
$IRAC1/H-IRAC1$ & 69,89,204,291,376,394,473,493,543,555,679,723\\
$IRAC1/K_S-IRAC1$ & 35,69,89,204,229,291,343,376,394,473,493,543,679,723\\
$IRAC1/WISE1-IRAC1$ & 22,89,181,229,343,364,394,449,493,543,592\\
$IRAC1/IRAC1-IRAC2$ & 291,592,679\\
$IRAC1/IRAC1-WISE2$ & 181,291,364,376,445,493,592,679,723\\
$IRAC1/IRAC1-IRAC3$ & 69,89,291,346,376,394,679,723\\
$IRAC1/IRAC1-IRAC4$ & 291,347,376,445,636,675,723\\
$IRAC1/IRAC1-WISE3$ & 291,376,445,592,649,672,679\\
$IRAC1/IRAC1-WISE4$ & 123,217,303,376,592,649,658,678,679,694,763\\
$IRAC1/IRAC1-MIPS24$ & 120,123,215,225,376\\
\midrule
\end{tabular}
\end{table*}

\clearpage

\begin{appendix}

\begin{figure*}
\center
\includegraphics[bb=125 360 490 720, scale=1.4]{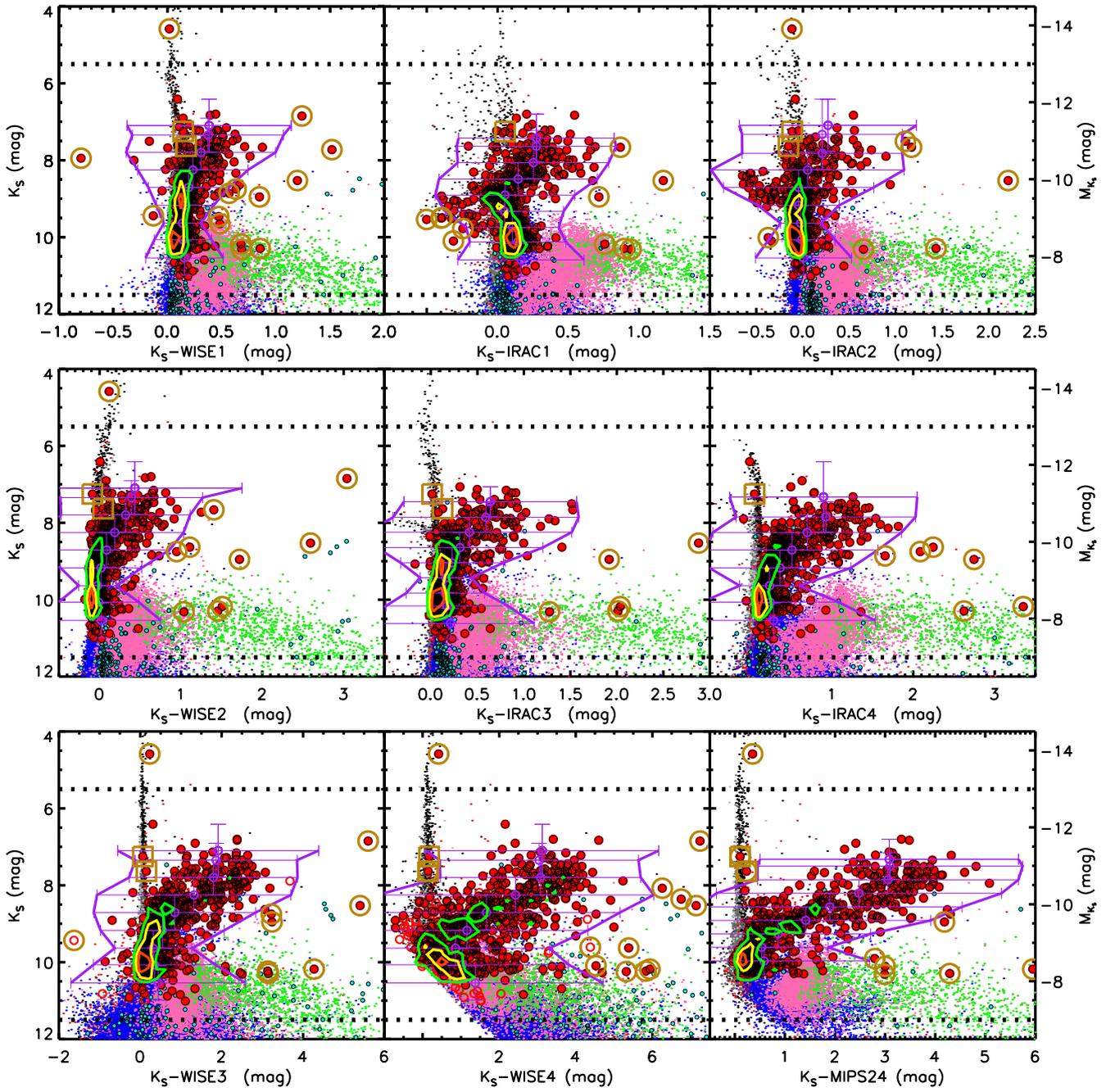}
\caption{Same as Fig~\ref{cmd_jhk} but for $K_S$ vs. $K_S-WISE1$, $K_S-IRAC1$, $K_S-IRAC2$, $K_S-WISE2$, $K_S-IRAC3$, $K_S-IRAC4$, $K_S-WISE3$, $K_S-WISE4$ and $K_S-MIPS24$. \label{ks_cmd}}
\end{figure*}

\begin{figure*}
\center
\includegraphics[bb=75 360 540 720, scale=1.1]{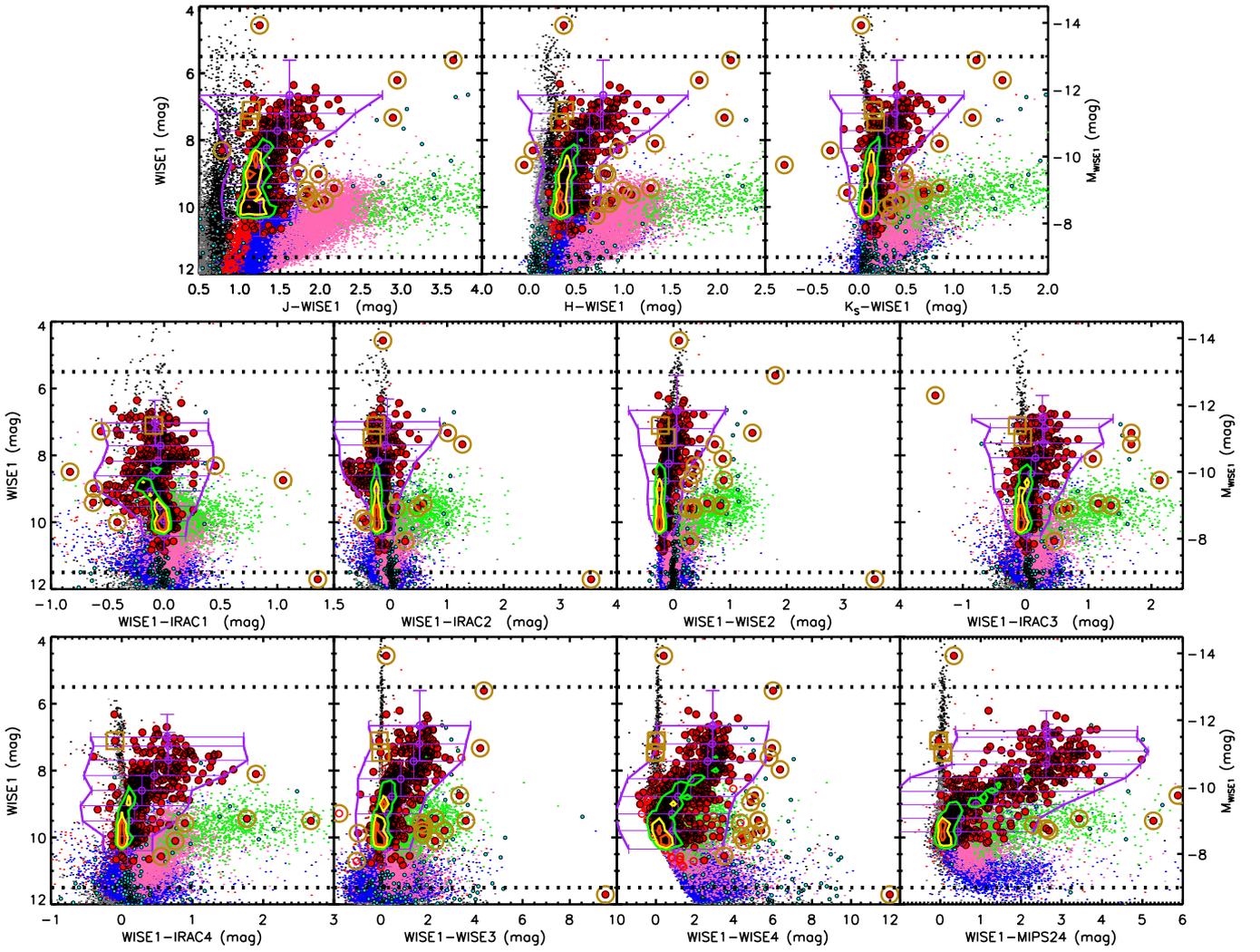}
\caption{Same as Fig~\ref{ks_cmd} but for WISE1 vs. $J-WISE1$, $H-WISE1$, $K_S-WISE1$, $WISE1-IRAC1$, $WISE1-IRAC2$, $WISE1-WISE2$, $WISE1-IRAC3$, $WISE1-IRAC4$, $WISE1-WISE3$, $WISE1-WISE4$ and $WISE1-MIPS24$. \label{w1_cmd}}
\end{figure*}

\begin{figure*}
\center
\includegraphics[bb=75 360 540 720, scale=1.1]{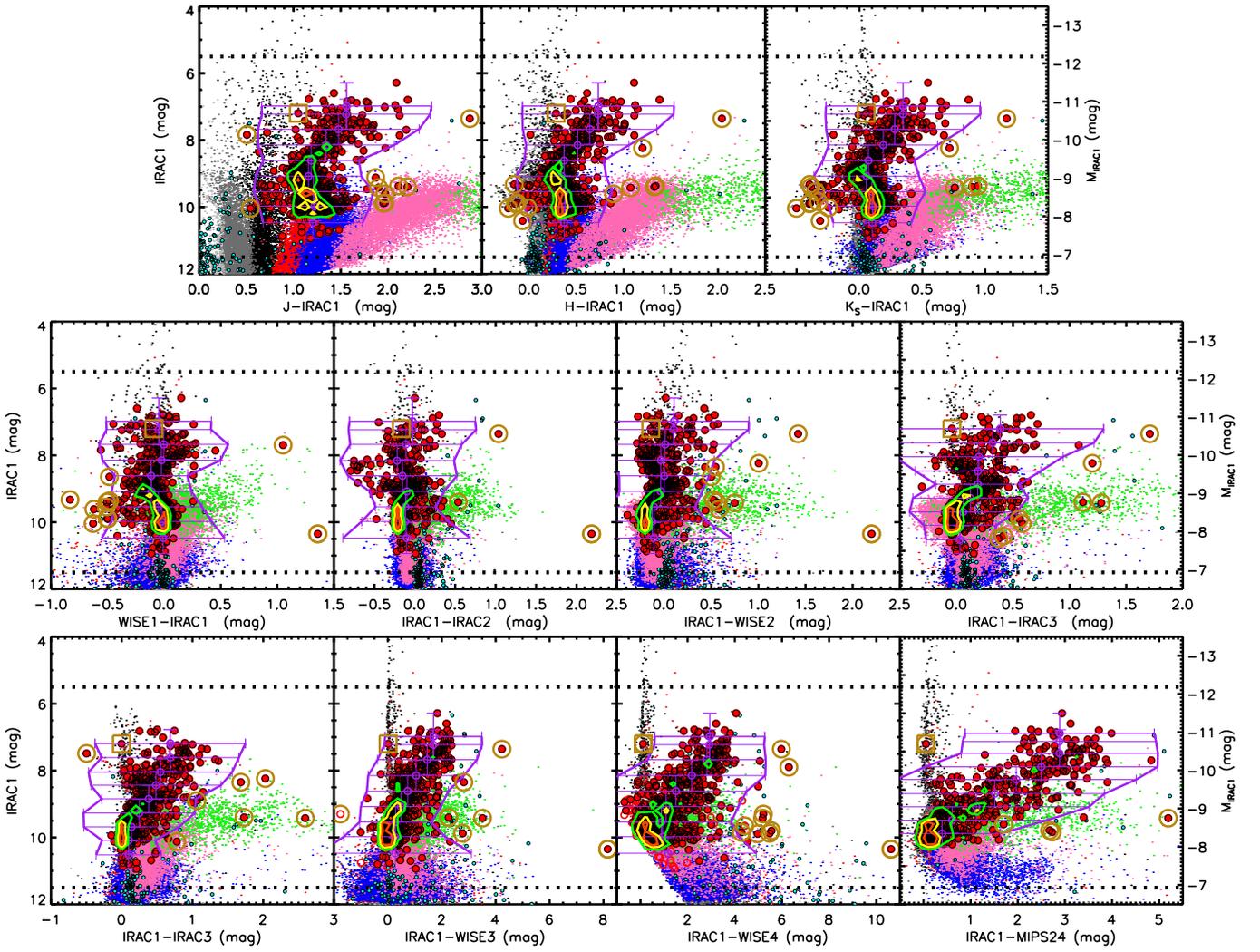}
\caption{Same as Fig~\ref{ks_cmd} but for IRAC1 vs. $J-IRAC1$, $H-IRAC1$, $K_S-IRAC1$, $WISE1-IRAC1$, $IRAC1-IRAC2$, $IRAC1-WISE2$, $IRAC1-IRAC3$, $IRAC1-IRAC4$, $IRAC1-WISE3$, $IRAC1-WISE4$ and $IRAC1-MIPS24$. \label{i1_cmd}}
\end{figure*}

\end{appendix}

\end{CJK*}

\end{document}